\documentclass[10pt,a4paper]{iopart}
\usepackage{iopams}  
\usepackage{graphicx,psfrag,bbm,latexsym,color,dcolumn,bm,dsfont,bbm,
color,mathrsfs,bbold,latexsym,amsfonts,amssymb}

\newcommand{\E}{\mbox{$\mathsf E$}}

\def\defi{{\buildrel \;def\; \over =}}

\begin{document}

\title[Quantum Phase Transitions induced by Infinite Dilution in the Fock Space]
{Quantum Phase Transitions induced by Infinite Dilution in the Fock Space: 
a General Mechanism. Proof and discussion.}

\author{Massimo Ostilli$^{1,2}$}

\address{$^1$\ Departamento de F{\'\i}sica da Universidade de Aveiro, 3810-193 Aveiro,
Portugal}
\address{$^2$\ Statistical Mechanics and Complexity Center (SMC), 
INFM-CNR SMC, Italy.}

\date{\today}

\begin{abstract}
We prove that lattice quantum systems may undergo
a first-order quantum phase transition through a general
mechanism which consists in an infinite
dilution of the states associated to (or, more in general, near to)
the lowest energy levels.
The equation giving the critical point is derived and discussed at several degrees of 
generalizations: given an infinitesimal portion of the Fock space, \textit{i.e.},
the cavity space, and its complement, \textit{i.e.}, the reservoir space,
in the thermodynamic limit, a first-order quantum phase transition takes place when 
the cavity and reservoir energies are equal.
A comparison with some particular exact solvable cases as, \textit{e.g.}, 
the Quantum Rem model, is made.
\end{abstract}


\maketitle

\section{Introduction}
Quantum Phase Transitions (QPT) are at the hearth of several fundamental phenomena like,
\textit{e.g.}, metal-insulator transition, superconductivity, superfluidity, etc..
\cite{Sach}.
QPT's can occur both in ordered and disordered models. 

In this paper we show that there exists a very general mechanism for which,
in the thermodynamic limit, and at zero temperature,
the system may undergo a first-order QPT through a infinite dilution in the Fock space.
Such a phenomena was conjectured by C. Presilla and the Author inspired by
what they found in a class of exact solvable random models \cite{OP5}. 
However, these random models are still very particular and do not allow to understand 
what could be the general mechanism leading to first-order QPT's piloted by an infinite dilution, if any.
 
Let us consider a lattice system of $N$ particles governed by an Hamiltonian operator $\hat{H}$ 
and let $\mathbb{F}=\{\bm{n}\}$ be the set of its states, \textit{i.e.}, the Fock space, and $H$ the associated
matrix representation. The matrix $H$ can always be splitted into two parts $H=K+V$,
where $V$ is a diagonal matrix while $K$ connects only different states.
The matrices $V$ and $K$ play the role of a potential and a kinetic operator in the Fock representation, respectively.
Let us indicate with $|E\rangle$ and $E$ the ground state (GS) and the ground state energy of the system, respectively.
Given an arbitrary configuration (or state) $\bar{\bm{n}}$, having potential $\bar{V}$, 
let us consider now the reduced Fock space $\widetilde{\mathbb{F}}=\mathbb{F}\setminus \bar{\bm{n}}$,
and let us indicate with $|\widetilde{E}\rangle$ and $\widetilde{E}$, 
the GS and GS energy of this reduced system, respectively.
Since $\widetilde{\mathbb{F}}$ and $\mathbb{F}$ differ for just one single configuration and 
the Fock dimension $|\mathbb{F}|$ grows rapidly with $N$, we expect that,
as the system size increases, $E$ and $\widetilde{E}$,  
will tend to differ for very small terms $\mathop{O}(N\bar{p})$, where $\bar{p}=1/|\mathbb{F}|$,
and $|E\rangle\sim |\widetilde{E}\rangle$.
In this paper we shall prove that 
this naive result is correct provided the potential energy $\bar{V}$, associated to the chosen 
state $\bar{\bm{n}}$, be bigger than the minimum possible potential $V_1$. 
In fact, if $\bar{V}=V_1$, 
when $\widetilde{E}$ is smaller than $V_1$, then - as one would expect - $|E-\widetilde{E}|/|E|=\mathop{O}(\bar{p})$ and
$\parallel |E\rangle-|\widetilde{E}\rangle\parallel^2=\mathop{O}(\bar{p})$, 
but when $\widetilde{E}$ is bigger than $V_1$, once approached
the configuration $\bar{\bm{n}}$ during a sufficiently long trajectory in the Fock space, 
the system will find energetically more convenient to remain in such a configuration for an infinite time.
We can see this by using an exact probabilistic representation (EPR) of the lattice quantum
dynamics \cite{BPDAJL}. Suppose for simplicity that the kinetic term $K$ has the
following form $K_{\bm{n},\bm{n}'}=-\eta \lambda_{\bm{n},\bm{n}'}$,
with $\eta>0$ and $\lambda_{\bm{n},\bm{n}'}=0,1$. Suppose also that $\bar{V}$ be non degenerate.
If we define the number of active links of the configurations $\bm{n}$ as 
$A(\bm{n})=\sum_{\bm{n}'}\lambda_{\bm{n},\bm{n}'}$ then, the EPR says that the matrix elements of the 
evolution operator $\exp(-Ht)$, $t$ being real, 
can be evaluated as expectations by 
\begin{eqnarray}
\label{I1}
\langle\bm{n}|e^{-Ht}|\bm{n}'\rangle=\E\left(\mathcal{M}^{[0,t)}\right),
\end{eqnarray}
where the stochastic functional $\mathcal{M}^{[0,t)}$ is defined as
\begin{eqnarray}
\label{I2}
\mathcal{M}^{[0,t)}\defi e^{\int_0^t [\eta A(\bm{n}_s)-V(\bm{n}_s)]ds},
\end{eqnarray}
and where $\bm{n}_s$ is the configuration reached at the time $s$ along trajectories 
extracted according to the Markov chain $P_{\bm{n},\bm{n}'}\propto \lambda_{\bm{n},\bm{n}'}$.
The steps in the trajectories extracted in the Fock space by the Markov chain take place 
at jump times Poissonian distributed with jumping rate $\eta$. 
This implies in particular that, once reached the configuration
$\bar{\bm{n}}$, the probability that
the system remains for a time $t$ in $\bar{\bm{n}}$, is $\exp(-\eta A(\bar{\bm{n}})t)$.
Given $\bar{\bm{n}}$, from Eqs. (\ref{I1}) and (\ref{I2}) we see therefore that in general,  
when $t\to\infty$, up to terms exponentially smaller in $t$, 
we can split the expectation and read the GS energy $E$ as follows
\begin{eqnarray}
\label{I3}
Ce^{-Et}\sim \widetilde{C}e^{-\widetilde{E}t}(1-\bar{p})+\bar{C}e^{-\bar{V}t}\bar{p},
\end{eqnarray}
where $C$, $\widetilde{C}$ and $\bar{C}$ are some finite constants.
Now, according to \cite{OP4}, it is always $E\leq V_1$, so that, for $t\to\infty$, the second term
in the above equation can be always neglected when $\bar{\bm{n}}$ is such that $\bar{V}>V_1$. 
If instead $\bar{V}=V_1$,
we see from Eq. (\ref{I3}) that the condition $\widetilde{E}/N=V_1/N$ for $N\to\infty$, possibly obtained 
by varying $\eta$ and the other Hamiltonian parameters entering in the matrix $V$, establishes
a critical point through which, in the thermodynamic limit, the system undergoes a first-order
QPT between a normal phase, in which $E/N\to\widetilde{E}/N$ and $|E\rangle\to|\widetilde{E}\rangle$,
and a frozen phase where instead $E/N\to V_1/N$ and $|E\rangle\to|\bar{\bm{n}}\rangle$.

In this paper we will see that, under the condition that $\bar{p}\to 0$ faster than $1/N$,
the above simple argument to derive Eq. (\ref{I3}) is essentially correct
and we can understand first-order QPT's as a result of an infinite dilution in the Fock space 
(not to be confused with a dilution of matter).
We will show that such a mechanism is actually much more general:
given a an infinitesimal portion of the Fock space, \textit{the cavity space},
and its complement, \textit{the reservoir space}, if - in the thermodynamic limit - 
the energies of the cavity and reservoir space are equal, the system undergoes a first-order QPT.

We think that this mechanism, due to its wide universality, 
could explain several already known QPT's as well as shed new light toward
many interesting less known phenomena. 

As an example in this paper we will show how the claimings of our theorem 
coincide with the case of the above mentioned random models \cite{OP5}, 
an exact solvable class of disordered models which includes, as a particular case,
the Quantum Rem Model \cite{BM}-\cite{Obuchi}, \textit{i.e.}, 
a model of disordered spins immersed in a transverse field. 

The main aim of this paper is however to prove the result
also in its most general form, postponing other possible tests and applications elsewhere. 
After stating the theorem, we will prove it at two different levels:
a simple approximate derivation (more suitable for the reader unfamiliar with stochastic processes)
for the non degenerate case, in which
finite size corrections are not calculated,
and then the exact derivation valid for the most general case
together with the finite size corrections.


\section{Theorem: QPT's as a result of an infinite dilution in the Fock space}
We now formalize the above phase transition scenario more explicitly.
Let $\mathbb{F}=\{\bm{n}\}$ be the Fock space of some system whose physical size (typically the number of particles)
we parametrize with $N$ and 
let $H$ be some matrix. In physical systems $M\defi |\mathbb{F}|$ in general grows with $N$ as a factorial,
however, for what follows, we do not need to specify any particular law for $M$. 
Let us split $H$ into two parts $H=K+V$,
where $V$ is a diagonal matrix while $K$ connects only different states.
Let us assume furthermore that the Markov chain associated to $K$ (see next section) be ergodic.
The matrices $V$ and $K$ play the role of a potential and a kinetic operator in the Fock representation, 
respectively \cite{NOTE}.
Finally, let us assume that, as occurs in physical systems, both the kinetic terms 
$K(\bm{n})\defi\sum_{\bm{n}'}K_{\bm{n},\bm{n}'}$ and the potential levels $V(\bm{n})\defi V_{\bm{n},\bm{n}}$ 
be extensive in the system size $N$ (at least for $N\to\infty$).

For pedagogical reasons, we find it convenient to separate the theorem in the following two parts.

\subsection{Cavity space associated to the potential level}
Let us order the different levels of $V$ as $(V_1<V_2<\ldots<V_m)$, $m$ being the number of different levels of $V$,
and let $(p_1,p_2,\ldots,p_m)$ be the associated density of the $V$'s normalized to 1: $\sum_{l=1}^mp_l=1$. 
In other words
\begin{eqnarray}
\label{THEO1} 
p_l = \frac{1}{M} \sum_{\bm{n}} \delta_{V_l,V{\bm{n}}} .
\end{eqnarray}
The number of different levels of $V$, $m$, may or may not grow with $N$.

Given a possible value of the potential $V_l$, with $1\leq l\leq m$,
let us consider the following reduced Fock spaces: 
\begin{eqnarray}
\label{THEO7} 
\mathbb{F}_l\defi\cup_{\bm{n}: V(\bm{n})=V_l}\bm{n},
\end{eqnarray}
which we shall call the \textit{cavity space},
and its complement 
\begin{eqnarray}
\label{THEO2} 
\widetilde{\mathbb{F}}\defi \mathbb{F}\setminus \cup_{\bm{n}: V(\bm{n})=V_l}\bm{n},
\end{eqnarray}
which we shall call the \textit{reservoir space}.
The matrix $H$ restricted to the spaces $\mathbb{F}_l$ and $\widetilde{\mathbb{F}}$
will have some GS and GS energy that we indicate with
$|E_l\rangle$ and $E_l$, and $|\widetilde{E}\rangle$ and $\widetilde{E}$, respectively.
Let us define as $e$, $\widetilde{e}$, $e_l$ and $v_l$, the energy density of $E$, $\widetilde{E}$, 
$E_l$, and $V_l$, in the thermodynamic limit, respectively 
(for what follows we will always assume the existence of these and other similar thermodynamic limits):
\begin{eqnarray}
\label{THEO3} 
e=\lim_{N\to\infty}\frac{E}{N}, \quad \widetilde{e}=\lim_{N\to\infty}\frac{\widetilde{E}}{N},
\quad e_l=\lim_{N\to\infty}\frac{E_l}{N}, 
\quad v_l=\lim_{N\to\infty}\frac{V_l}{N}.
\end{eqnarray}

\subsubsection{Finite degenerate case}
Let us suppose first that the level $V_l$ be finitely degenerate in the limit $N\to\infty$. 
Note that this implies in particular that be $p_l\to 0$ for $N\to\infty$. 
Let be $p_lN\to 0$, then for $l>1$
\begin{eqnarray}
\label{THEO4} 
e=\widetilde{e},  
\end{eqnarray}
while for $l=1$
\begin{eqnarray}
\label{THEO5} 
e=\left\{
\begin{array}{l}
\widetilde{e}, \quad \widetilde{e}\leq v_1, \\
v_1, \quad \widetilde{e}> v_1.
\end{array}
\right.
\end{eqnarray}
Eqs. (\ref{THEO4}) and (\ref{THEO5}) tell us that, in the thermodynamic limit, 
the GS of the reduced system approaches the GS of the full system if $l>1$, while if $l=1$, we have 
that the equation 
\begin{eqnarray}
\label{THEO6} 
\widetilde{e}=v_1,
\end{eqnarray}
as an equation in the space of the ``Hamiltonian'' parameters entering $H$,
represents a critical condition where a first-order QPT takes place: we will prove that,
for $\widetilde{e}\leq v_1$ - as in the normal case $l>1$ - 
the GS's $|E\rangle$ and $|\widetilde{E}\rangle$ differ for terms $\mathop{O}(p_1)^{1/2}$, 
and $|E-\widetilde{E}|=\mathop{O}(Np_1)$, while
for $\widetilde{e}> v_1$ the system gets frozen and $|E\rangle$ tends to $\bm{n}_1$,
the configuration associated to the lowest potential level $V_1$.

\subsubsection{Infinite degenerate case} 
Let us consider now a more general case in which the level $V_l$ may be infinitely degenerate in the limit $N\to\infty$. 
Now, the cavity space associated to the level $V_l$ has, not only an extensive potential
energy, but also an extensive ``kinetic energy''. 
Let the states in $\mathbb{F}_l$ be connected. Then, if for $N\to\infty$, $p_lN \to 0$, we have
\begin{eqnarray}
\label{THEO8} 
e=\left\{
\begin{array}{l}
\widetilde{e}, \quad \widetilde{e}\leq e_l, \\
e_l, \quad \widetilde{e}> e_l.
\end{array}
\right.
\end{eqnarray}
For $l=1$, Eq. (\ref{THEO8}) is analogous to Eq. (\ref{THEO5}) and represents the natural generalization 
of the former case to the infinitely degenerate case, the critical condition now being 
\begin{eqnarray}
\label{THEO9} 
\widetilde{e}=e_1.
\end{eqnarray}
Note however, that unlike the former case, 
we cannot assume the analogous of Eq. (\ref{THEO4}) for $l>1$, and in particular
here we are not able to know a priori whether a more general critical condition 
$\widetilde{e}=e_l$ may or may not have solutions. However, whatever be $l$, according
to Eq. (\ref{THEO8}), if the equation $\widetilde{e}=e_l$ admits solutions, 
the system will undergo a first-order QPT phase transition with the
same mechanism as in the finite degenerate case with a normal phase and a frozen
one where the GS lives in the Fock space $\mathbb{F}_l$.

\subsection{General case} 
Finally, we have the following natural generalization.
Let $\bar{\mathbb{F}}$ be any connected subspace of $\mathbb{F}$ having a fraction
$\bar{p}$ of configurations with $\bar{p}N\to 0$ for $N\to\infty$ \cite{NOTE0}, then
in the thermodynamic limit, for the energy density $e$ we have 
\begin{eqnarray}
\label{THEO10} 
e=\left\{
\begin{array}{l}
\widetilde{e}, \quad \widetilde{e}\leq \bar{e}, \\
\bar{e}, \quad \widetilde{e}> \bar{e},
\end{array}
\right.
\end{eqnarray}
where the density energies $\widetilde{e}$ and $\bar{e}$ are those associated
to the spaces $\widetilde{\mathbb{F}}$ and $\bar{\mathbb{F}}$, respectively.
As in the previous case, we are not able to say nothing about
the possibility that the equation $\widetilde{e}=\bar{e}$ has or has not solutions
in the space of the Hamiltonian parameters present in the original matrix $H$.
Again, in the affirmative case, we have a first-order QPT 
taking place with the same mechanism.

Concerning the finite system in which then $\bar{p}$ is small but finite,
we will also prove that, near the critical point, the difference between
$E$ and $\bar{E}$, or $\widetilde{E}$, is of the order $\mathop{O}(\bar{\pi}K_{\mathrm{out}})$,
$K_{\mathrm{out}}$ being the kinetic energy by which, once entered in the cavity, with
asymptotic probability $\bar{\pi}$,
the system is expelled out from
the cavity space itself (\textit{i.e.}, the space $\bar{\mathbb{F}}$). More precisely, $\bar{\pi}$ is 
the invariant measure of the Markov chain associated to the matrix elements of the ``kinetic'' matrix $K$.
Note that, as we will point out later, since under the hypothesis of the theorem the matrix $K$ is connected
and the kinetic terms extensive, 
it is always $\bar{\pi}\leq \bar{p}$, for $N\to\infty$, so that
the condition $\bar{p} N\to 0$ ensures also that be $\bar{\pi}N\to 0$ \cite{NOTE0}.
The detailed expression for $K_{\mathrm{out}}$ is quite cumbersome and is left in
the proof (see Eq. (\ref{QLemma16})). The energy $K_{\mathrm{out}}$, besides to be dependent on
the kinetic terms associated to the boundary of the cavity, depends also on the overlaps that the reservoir and cavity GS's 
have on their internal boundaries $\partial\widetilde{\mathbb{F}}$ and $\partial\bar{\mathbb{F}}$, receptively. 
The larger are these Fock-boundaries contributions, the stronger will be the finite size effects.

\section{Proof}
\label{proof}
\subsection{Exact probabilistic representation (EPR)}
The derivation of Eqs. (\ref{THEO4}), (\ref{THEO5}) and (\ref{THEO10}) 
follows from an exact probabilistic representation derived in \cite{BPDAJL} and that
here we recall briefly. 
In general, given an arbitrary Hamiltonian matrix ${H}$, 
each row $\bm{n}$ of the corresponding hopping matrix ${K}$ 
has a different number of non zero elements, let say $A(\bm{n})$. 
We call $A(\bm{n})$ the number of active links of the state $\bm{n}$ (in graph theory $A(\bm{n})$ is called 
degree of the vertex $\bm{n}$).
Given the matrix ${H}$, and its hopping and potential matrices, $K$ (the non diagonal part of $H$) 
and ${V}$ (the diagonal part of $H$), 
respectively, we define a virtual dynamics as follows.
Let us parametrize the matrix ${K}$ as
\begin{eqnarray}
\label{Kf} 
K_{\bm{n},\bm{n}'}=-\lambda_{\bm{n},\bm{n}'} ~\eta_{\bm{n},\bm{n}'}, 
\end{eqnarray}
such that $|\lambda_{\bm{n},\bm{n}'}|$ can be either 0 or 1  
and $\eta_{\bm{n},\bm{n}'}>0$.
In graph theory the matrix with elements $|\lambda_{\bm{n},\bm{n}'}|$ is
known as the adjacency matrix: in fact, it establishes whether
two given states $\bm{n},\bm{n}'$ are first neighbors or not.
We consider the Markov chain defined by
the transition matrix ${P}$ with elements
\begin{eqnarray}
\label{Pf} 
P_{\bm{n},\bm{n}'}=\frac{|K_{\bm{n},\bm{n}'}|}{R(\bm{n})},
\end{eqnarray}
where 
\begin{eqnarray}
\label{Af} 
R(\bm{n}) = \sum_{\bm{n}'} |K_{\bm{n},\bm{n}'}| .
\end{eqnarray}
Starting from a given initial configuration $\bm{n}_0$, 
we draw a new configuration $\bm{n}_1$ with probability 
$P_{\bm{n}_0,\bm{n}_1}$.
By iterating this procedure for $N$ steps 
we construct a path, or trajectory, in the space $\mathbb{F}$
$\bm{n}_0,\bm{n}_1,\ldots,\bm{n}_N$. 
Note that, since $P_{\bm{n},\bm{n}'}$ is supposed to be ergodic,
after an infinite number of jumps the Markov chain converges
to the invariant measure $\bm{\pi}$, which coincides with 
the left eigenvector of the transition matrix, 
$\bm{\pi}^{\mathrm{T}} {P} = \bm{\pi}^{\mathrm{T}}$. 
It is simple to verify that \cite{OP4}
\begin{eqnarray}
\label{PI} 
\pi(\bm{n})=\frac{R(\bm{n})}{\sum_{\bm{n}'}R(\bm{n}')}.
\end{eqnarray}
Notice that, the hypothesis that the Markov chain be ergodic, implies in particular
that for any $\bm{n}$ be $R(\bm{n})>0$, so that from Eq. (\ref{PI}) we have
\begin{eqnarray}
\label{PIb} 
\pi(\bm{n})\leq \frac{R(\bm{n})}{\mathrm{min}_{\bm{n}'\in \mathbb{F}}R(\bm{n}')}\frac{1}{|\mathbb{F}|},
\end{eqnarray}
and more in general, if $\bar{\mathbb{F}}$ is a subspace with a measure $\bar{p}$ 
coming from Eq. (\ref{THEO1}), we have 
\begin{eqnarray}
\label{PIc} 
\bar{\pi}\defi \pi(\bar{\mathbb{F}})=\frac{\sum_{\bm{n}\in\bar{\mathbb{F}}}R(\bm{n})}{\sum_{\bm{n}}R(\bm{n})}
\leq \frac{\mathrm{max}_{\bm{n}\in \bar{\mathbb{F}}}R(\bm{n})}
{\mathrm{min}_{\bm{n}\in \mathbb{F}}R(\bm{n})}\bar{p},
\end{eqnarray}
so that, if the $R$'s are all extensive, and if, in the thermodynamic limit, 
$\bar{p}\to 0$, also $\bar{\pi}\to 0$ (see note \cite{NOTE0}).

Along each finite path with $N$ steps we have the sequences of data
$A_0,A_1,\ldots,A_N$, $R_0,R_1,\ldots,R_N$, $V_0,V_1,\ldots,V_N$,
$\lambda_1,\ldots,\lambda_N$, 
and $\eta_1,\ldots,\eta_N$, where
\begin{eqnarray}
\label{Lf} 
A_k=A(\bm{n}_k), \qquad k=0,\dots,N,
\\
R_k=R(\bm{n}_k), \qquad k=0,\dots,N,
\\
V_k=V(\bm{n}_k), \qquad k=0,\dots,N,
\\
\lambda_k=\lambda_{\bm{n}_{k-1},\bm{n}_k}, \qquad k=1,\dots,N,
\\
\eta_k=\eta_{\bm{n}_{k-1},\bm{n}_k}, \qquad k=1,\dots,N.
\end{eqnarray} 
Note that, by definition, along a path of the Markov chain $|\lambda_k|\equiv 1$.

If at each step of the Markov chain we extract randomly a jump time $s$ distributed according 
to a Poisson process of parameter $\rho$, and along each trajectory 
we indicate the sequence of these drawn jump times as $s_1,s_2,\ldots,s_{N_t}$, $N_t$
being the number of jumps occurred along the trajectory,
then the following exact probabilistic representation holds~\cite{BPDAJL}
\begin{eqnarray}
\label{TheFormulaa}
\langle \bm{n}|e^{-Ht} | \bm{n}_0\rangle &=&  
\E  \left( \delta_{ \bm{n} , \bm{n}_{N_t}} 
{\cal M}^{[0,t)}_{\bm{n}_0} \right), 
\end{eqnarray}
where the stochastic functional ${\cal M}^{[0,t)}_{\bm{n}_0}$ is defined as
\begin{eqnarray}
\label{FORMULA D}
{\cal M}^{[0,t)}_{\bm{n}_0} =
\left( 
\prod_{k=1}^{N_{t}} \frac{\eta_k}{\rho} \lambda_k
e^{\left[\rho A_{k-1}-V_{k-1}\right](s_{k}-s_{k-1})} 
\right)  
e^{\left[\rho A_{N_{t}}-V_{N_{t}}\right](t-s_{N_{t}})}
\end{eqnarray}
if $N_t > 0$, and 
${\cal M}^{[0,t)}_{\bm{n}_0} = e^{(\rho A_0-V_0)t}$ if $N_t=0$,
and the symbol $\E(\cdot)$ means expectation over the Poisson processes and the Markov chain.
In Eq.~(\ref{FORMULA D}), we put $s_0=0$. 
According to this representation, we have in particular 
\begin{eqnarray}
\label{TheFormulab}
\sum_{\bm{n}} \langle \bm{n}|e^{-Ht} | \bm{n}_0\rangle =  
\E  \left( {\cal M}^{[0,t)}_{\bm{n}_0} \right),  
\end{eqnarray}
so that we can evaluate the ground-state energy as
\begin{eqnarray}
\label{E0} 
E_0 
= \lim_{t \to \infty} -\partial_t \log 
\E \left( \mathcal{M}^{[t,0)}_{\bm{n}_0} \right).
\end{eqnarray}
In the following we rewrite Eq. (\ref{FORMULA D}) in a more compact form as
\begin{eqnarray}
\label{EPR}
{\cal M}^{[0,t)}_{\bm{n}_0} =
\left( 
e^{\int_0^t\left[\rho A\left(\bm{n}_s\right)-V\left(\bm{n}_s\right)\right]ds} 
\prod_{k=1}^{N_{t}} \frac{\eta_k}{\rho} \lambda_k
\right).  
\end{eqnarray}
More in general, if at each active link $(\bm{n},\bm{n}')$ we associate a Poisson process 
with rate $\rho_{\bm{n},\bm{n}'}$ dependent on the
link, the following generalization holds:
\begin{eqnarray}
\label{EPRG}
{\cal M}^{[0,t)}_{\bm{n}_0} =
\left( 
e^{\int_0^t\left[R\left(\bm{n}_s\right)-V\left(\bm{n}_s\right)\right]ds} 
\prod_{k=1}^{N_{t}} \frac{\eta_k}{\rho_k} \lambda_k
\right),
\end{eqnarray}
where $R(\bm{n})$, the weighted number of active links, is defined in Eq. (\ref{Af}). 
For what follows we will always make the choice
\begin{eqnarray}
\label{CHOICE}
\rho_{\bm{n},\bm{n}'}=\eta_{\bm{n},\bm{n}'}
\end{eqnarray}
so that the product in Eq. (\ref{CHOICE}) simplifies in
\begin{eqnarray}
\label{EPRGC}
{\cal M}^{[0,t)}_{\bm{n}_0} =
e^{\int_0^t\left[R\left(\bm{n}_s\right)-V\left(\bm{n}_s\right)\right]ds} 
\prod_{k=1}^{N_{t}} \lambda_k.
\end{eqnarray}

\subsection{Proof of Eqs. (\ref{THEO4})-(\ref{THEO5}) (simple derivation)}
Let us consider the degenerate case, \textit{i.e.}, the case in which the cavity
consists of a single state $\bar{\bm{n}}$.
In Sec. II we have indicated with $\widetilde{E}$ the GS energy of the system
living in the reduced Fock space $\widetilde{\mathbb{F}}=\mathbb{F}\setminus \bar{\bm{n}}$
and having (for $\bm{n}\neq \bar{\bm{n}}$) the same matrix Hamiltonian representation $H$
(alternatively, such a system could be still seen as a system living in the original Fock space 
$\mathbb{F}$ having as operators $\widetilde{{K}}={K}$  
and $\widetilde{{V}}_{\bm{n},\bm{n}}={V}_{\bm{n},\bm{n}}$ for $\bm{n}\neq \bar{\bm{n}}$
and $\widetilde{{V}}_{\bar{\bm{n}},\bar{\bm{n}}}=+\infty$). This Hamiltonian in the following
will be indicated by $\widetilde{H}$. 
We find it convenient to formally define also an Hamiltonian  
associated to the single state $\bar{\bm{n}}$ and taking (for the moment being) the single value 
$\bar{H}=\bar{V}$.

Given an initial configuration $\bm{n}_0\in \widetilde{\mathbb{F}}$ \cite{NOTE2},
let $k_t$ be the stochastic variable counting how many times the system, starting from $\bm{n}_0$, enters in the
state $\bar{\bm{n}}$ during the time interval $[0,t)$. We can decompose the expectation as
\begin{eqnarray}
\label{Q1}
\E\left(\mathcal{M}^{[0,t)}\right)=\sum_{k=0}^{\infty}\E\left(\mathcal{M}^{[0,t)}|k_t=k\right)Q_t(k),
\end{eqnarray}
where in the rhs we have introduced $Q_t(k)$, the probability that the system enters in the state  
$\bar{\bm{n}}$ exactly $k$ times during the time interval $[0,t)$ (dependencies on the initial configuration
are understood).  
Let $t_1,t_2,\ldots,t_k$ be the jump times at which the system enters in the state $\bar{\bm{n}}$
through the jump of some active link having signs 
$\lambda_{\bm{n}_{t_1^-},\bm{n}_{t_1^+}},\ldots,\lambda_{\bm{n}_{t_k^-},\bm{n}_{t_k^+}}$, respectively, and 
$\tau_1,\tau_2,\ldots,\tau_k$, the corresponding living times in $\bar{\bm{n}}$,
from which the system leaves through a further jump of some active link having sign
$\lambda_{\bm{n}_{t_1+\tau_1^-},\bm{n}_{t_1+\tau_1^+}},\ldots,\lambda_{\bm{n}_{t_k+\tau_k^-},\bm{n}_{t_k+\tau_k^+}}$, respectively.
Here, to avoid confusion, we have introduced the symbols for the configurations
immediately before the jump taking place at the time $t_k$, as $\bm{n}_{t_k^-}$, 
and immediately after the jump as $\bm{n}_{t_k^+}$,
and similarly for $\bm{n}_{t_k+\tau_k^-}$ and $\bm{n}_{t_k+\tau_k^+}$.  
Then, from Eq. (\ref{EPRGC}) we see that
Eq. (\ref{Q1}) becomes
\begin{eqnarray}
\label{Q2}\fl
\E\left(\mathcal{M}^{[0,t)}\right)=\sum_{k=0}^{\infty}\E\left(
\widetilde{\mathcal{M}}^{[0,t_1)}\lambda_{\bm{n}_{t_1^-},\bm{n}_{t_1^+}}\bar{\mathcal{M}}^{[t_1,t_1+\tau_1)}
\lambda_{\bm{n}_{t_1+\tau_1^-},\bm{n}_{t_1+\tau_1^+}}
\cdots \right. \nonumber \\
\fl
\left. \times \widetilde{\mathcal{M}}^{[t_{k-1}+\tau_{k-1},t_k)}
\lambda_{\bm{n}_{t_k^-},\bm{n}_{t_k^+}}
\bar{\mathcal{M}}^{[t_k,t_k+\tau_k)}
\lambda_{\bm{n}_{t_k+\tau_k^-},\bm{n}_{t_k+\tau_k^+}}
\widetilde{\mathcal{M}}^{[t_k+\tau_k,t)}
|k_t=k\right)Q_t(k),
\end{eqnarray}
where the stochastic functionals
$\widetilde{\mathcal{M}}^{[0,t)}$ and $\bar{\mathcal{M}}^{[0,t)}$
are defined similarly to the functional ${\mathcal{M}}^{[0,t)}$ introduced at the beginning but
the Hamiltonian $H$ is replaced with the Hamiltonian's $\widetilde{H}$ and $\bar{H}$, respectively.
Let us note that the two set of stochastic variables $t_1,t_2,\ldots,t_k$ and $\tau_1,\tau_2,\ldots,\tau_k$ are independent.
Furthermore, the variables $\tau_1,\tau_2,\ldots,\tau_k$ are almost each other independent, the only constraint being that
their sum cannot overcome $t$. So that, taking into account that once entered in
$\bar{\bm{n}}$, the infinitesimal probability that the system leaves $\bar{\bm{n}}$ 
after a time between $\tau$ and $\tau+d\tau$
is $d\tau\bar{R}\exp(-\bar{R}\tau)$, with $\bar{R}\defi R(\bar{\bm{n}})$ \cite{NOTE1}, 
according to Eq. (\ref{EPRGC}) we have
\begin{eqnarray}
\label{Q3}\fl
\E\left(\mathcal{M}^{[0,t)}\right)=\sum_{k=0}^{\infty}\int_{0}^{'t} dx_1dx_2\ldots dx_k
\E\left(\widetilde{\mathcal{M}}^{[0,t_1)}\widetilde{\mathcal{M}}^{[t_1+x_1,t_2)}\cdots 
\widetilde{\mathcal{M}}^{[t_k+x_k,t)} \right. \nonumber \\
\fl
\left. \times 
e^{-\bar{V}(x_1+x_2\ldots+x_k)}|k_t=k; \tau_1=x_1,\ldots, \tau_k=x_k\right)
\bar{R}^k \left\langle\prod_{l=1}^k \lambda_{t_l}\lambda_{t_l+x_l}\right\rangle_tQ_k(t),
\end{eqnarray}
where $\int_{0}^{'t} dx_1dx_2\ldots dx_k$ is a shorthand notation for the integral
extended to the only $k$-dimensional set where the variables satisfy the constraint $0\leq x_1+x_2+\ldots +x_k\leq t$,
and we have introduced the following stochastic normalized average 
\begin{eqnarray}
\label{Q3b}\fl
\left\langle\prod_{l=1}^k \lambda_{t_l}\lambda_{t_l+x_l}\right\rangle_t\defi\nonumber \\
\fl\frac{
\E\left(\widetilde{\mathcal{M}}^{[0,t_1)}\lambda_{t_1}\lambda_{t_1+x_1}\widetilde{\mathcal{M}}^{[t_1+x_1,t_2)}
\lambda_{t_2}\lambda_{t_2+x_2}\cdots 
\lambda_{t_k}\lambda_{t_k+x_k}\widetilde{\mathcal{M}}^{[t_k+x_k,t)} |k_t=k\right)
}
{\E\left(\widetilde{\mathcal{M}}^{[0,t_1)}\widetilde{\mathcal{M}}^{[t_1+x_1,t_2)}\cdots 
\widetilde{\mathcal{M}}^{[t_k+x_k,t)} |k_t=k\right).
}
\end{eqnarray}
Note that, in the above expression, we have for brevity written the expectations appearing in the
numerator and denominator as non conditioned by the times $\tau_1=x_1,\ldots,\tau_k=x_k$. 
However, in the limit $t\to\infty$, the stochastic normalized average, which is the ratio of the expectations,
becomes a constant independent on the $\tau$'s.

We observe that, in each trajectory, the total time $\widetilde{t}$ 
during which the system is not in the state $\bar{\bm{n}}$ is
$\widetilde{t}=t-(x_1+\ldots +x_k)$. Furthermore, for any fixed value of $k$, 
each interval, $[t_{l}+x_l,t_{l+1})$, in which the system is not in the state $\bar{\bm{n}}$, has a length 
which is of the order $\mathop{O}(\widetilde{t}/k)$. Therefore, when $t\to\infty$, 
the stochastic functionals $\widetilde{\mathcal{M}}^{[t_l+x_l,t_{l+1})}$ in Eq. (\ref{Q3}) get uncorrelated and
for any finite $k$ we are left with \cite{NOTE3}
\begin{eqnarray}
\label{Q4}\fl
&&\E\left(
\widetilde{\mathcal{M}}^{[0,t_1)}\widetilde{\mathcal{M}}^{[t_1+x_1,t_2)}\cdots 
\widetilde{\mathcal{M}}^{[t_k+x_k,t)} 
|k_t=k; \tau_1=x_1,\ldots, \tau_k=x_k\right)\nonumber \\
&&= C_t(k)e^{-\widetilde{E}[t-(x_1+\ldots+x_k)]},
\end{eqnarray}
where $\widetilde{E}$ is the GS energy of the system with Hamiltonian $\widetilde{H}$,
and $C_t(k)$ is a bounded quantity smoothly dependent on $t$.
In the proof for the general case we will show that both these constants and
the stochastic averages (\ref{Q3b}) are well bounded functions. 
More precisely we will show that the net result is given by 
\begin{eqnarray}
\label{Q3c}\fl
C_t(k)\bar{R}^k\left\langle\prod_{l=1}^k \lambda_{t_l}\lambda_{t_l+x_l}\right\rangle_t=
\widetilde{C}_{\bm{n}_0}\left(-K_{\mathrm{out}}\right)^k
\left(1+\mathop{O}\left(e^{-(\widetilde{E}_1-\widetilde{E})t}\right)\right),
\end{eqnarray}
where: $\widetilde{E}_1>\widetilde{E}$ is the energy of the first excited state of the reservoir, 
$\widetilde{C}_{\bm{n}_0}=\mathop{O}(1)$ is a constant depending only on the initial condition $\bm{n}_0$,
and $K_{\mathrm{out}}$,  
has the meaning of a kinetic energy pushing the system out of the 
configuration $\bar{\bm{n}}$.
It is in fact $K_{\mathrm{out}}\sim -\left(\bar{R}^+-\bar{R}^-\right)$,
where $\bar{R}^+$ and $\bar{R}^-$ are the kinetic terms obtained as
sums over the positive ($\lambda=1$) and negative ($\lambda=-1$) active links, respectively.

By using now Eqs. (\ref{Q4}) and (\ref{Q3c}) in Eq. (\ref{Q3}), up to terms exponentially smaller in $t$, we get
\begin{eqnarray}
\label{Q5}\fl
\E\left(\mathcal{M}^{[0,t)}\right)&=& \widetilde{C}_{\bm{n}_0}e^{-\widetilde{E}t}
\sum_{k=0}^{\infty}\left(-K_{\mathrm{out}}\right)^k 
\int_{0}^{'t} dx_1dx_2\ldots dx_k
e^{(\widetilde{E}-\bar{V})(x_1+\ldots+x_k)}Q_t(k).
\end{eqnarray}

We have now to understand the behavior of $Q_t(k)$. We recall that the total number of jumps
made by the system during the interval $[0,t)$ is of the order $\langle R\rangle t$,
where $\langle R\rangle$ 
is the average of the weighted number of active links $R$ 
along a trajectory making an infinite number of jumps \cite{NOTE4}.
Now, the way the system jumps from one configuration $\bm{n}$ to the next one $\bm{n}'$ 
is a finite ergodic Markov chain $P_{\bm{n},\bm{n}'}$. 
Therefore, there exists a critical
value $N_c$ for which, if the number of jumps $N$ satisfies $N\gg N_c$,
up to terms exponentially smaller in $N$ one has $(P^N)_{\bm{n},\bm{n}'}=\pi(\bm{n}')$, where $\bm{\pi}$ is
the invariant measure given by Eq. (\ref{PI}). 
In particular, this implies that, for $N\gg kN_c$, up to terms exponentially smaller in $N$, 
the probability $\mathcal{P}_N(k)$ that the Markov chain visits $k$ times the state $\bm{n}$ 
is $\mathcal{P}_N(k)=(\pi(\bm{n}))^k$. 
Therefore, there exists a critical value $t_c\sim N_c/\langle R\rangle$ for which, 
if $t\gg kt_c$, up to terms exponentially smaller in $t$, we have also 
$Q_t(k)=(\pi(\bar{\bm{n}}))^k$ \cite{NOTE5}.
Furthermore, 
recalling that $0\leq Q_t(k)\leq 1$, and that
$\int_0^{'t} dx_1\ldots dx_k =t^k/k!$, 
we see immediately that the series in the rhs of Eq. (\ref{Q5})
is absolutely convergent for any $t$. This fact allows us to pass the limit of $t$ large under the series symbol
and to analyze the behavior of the system by simply analyzing term by term the single contributions of the series.
The integrals involved in the rhs of Eq. (\ref{Q5}) cannot be expressed in a closed form.
However, the analysis of the first few already makes clear the general structures. 
For $\bar{V}\neq \widetilde{E}$ we have:
\begin{eqnarray}
\label{Q6}\fl
\E\left(\mathcal{M}^{[0,t)}\right)&=& e^{-\widetilde{E}t}\widetilde{C}_{\bm{n}_0}Q_t(0)+
\left(e^{-\bar{V}t}-e^{-\widetilde{E}t}\right)\frac{1}{\widetilde{E}-\bar{V}}\widetilde{C}_{\bm{n}_0}
\left(-K_{\mathrm{out}}\right)Q_t(1)\nonumber \\
&& \fl + \left[e^{-\bar{V}t}t\frac{1}{\widetilde{E}-\bar{V}}-\left(e^{-\widetilde{V}t}-e^{-\bar{E}t}\right)
\frac{1}{\left(\widetilde{E}-\bar{V}\right)^2}
\right]\widetilde{C}_{\bm{n}_0}\left(-K_{\mathrm{out}}\right)^2 Q_t(2) \nonumber \\
&& +\ldots + \widetilde{C}_{\bm{n}_0}\mathcal{G}_k\left(t\right)\left(-K_{\mathrm{out}}\right)^k Q_t(k)+\ldots,
\end{eqnarray}
where $\mathcal{G}_k\left(t\right)$ is a multivariate polynomial 
in the variables $\exp(-\bar{V}t)$, $\exp(-\widetilde{E}t)$, and $t,t^2\ldots,t^{k-1}$.
Let $\bar{k}_t$ be an integer such that $1\leq \bar{k}_t\ll t/t_c$. 
By using $Q_t(k)\simeq \pi(\bar{\bm{n}})^k$, for $1\leq k\leq \bar{k}_t$, and then 
$Q_t(0)\simeq 1-\sum_{k=1}^\infty \pi(\bar{\bm{n}})^k$ for $t$ large enough, up to terms exponentially smaller in $t$,
Eq. (\ref{Q6}) becomes
\begin{eqnarray}
\label{Q8}\fl
\E\left(\mathcal{M}^{[0,t)}\right)= e^{-\widetilde{E}t}\frac{1-2\pi(\bar{\bm{n}})}{1-\pi(\bar{\bm{n}})}
\widetilde{C}_{\bm{n}_0}+
\left(e^{-\bar{V}t}-e^{-\widetilde{E}t}\right)\frac{1}{\widetilde{E}-\bar{V}}\widetilde{C}_{\bm{n}_0}
\left(-K_{\mathrm{out}}\right)\pi(\bar{\bm{n}})\nonumber \\
\fl + \left[e^{-\bar{V}t}t\frac{1}{\widetilde{E}-\bar{V}}-\left(e^{-\widetilde{V}t}-e^{-\bar{E}t}\right)
\frac{1}{\left(\widetilde{E}-\bar{V}\right)^2}
\right]\widetilde{C}_{\bm{n}_0}\left(-K_{\mathrm{out}}\pi(\bar{\bm{n}})\right)^2\nonumber \\
\fl +\ldots+\widetilde{C}_{\bm{n}_0}\mathcal{G}_k\left(t\right)
\left(-K_{\mathrm{out}}\pi(\bar{\bm{n}})\right)^k+\ldots 
+\widetilde{C}_{\bm{n}_0}\mathcal{G}_{\bar{k}_t}\left(t\right)
\left(-K_{\mathrm{out}}\pi(\bar{\bm{n}})\right)^{\bar{k}_t}
+\mathcal{R}_t,
\end{eqnarray} 
where $\mathcal{R}_t$ is the rest of the series taking into account the contributions for 
$k>\bar{k}_t$. Note that, by construction, for $t\to\infty$, $\bar{k}_t\to\infty$, 
$\mathcal{R}_t\to 0$ and Eq. (\ref{Q8}) becomes exact.

Concerning the case $\bar{V}=\widetilde{E}$ instead we have
\begin{eqnarray}
\label{Q9}\fl
\E\left(\mathcal{M}^{[0,t)}\right)= e^{-\bar{V}t}\left(\frac{1-2\pi(\bar{\bm{n}})}{1-\pi(\bar{\bm{n}})}
\widetilde{C}_{\bm{n}_0}+
t\widetilde{C}_{\bm{n}_0}\left(-K_{\mathrm{out}}\right)\pi(\bar{\bm{n}})+\frac{t^2}{2}
\widetilde{C}_{\bm{n}_0}\left(-K_{\mathrm{out}}\pi(\bar{\bm{n}})\right)^2\right. \nonumber \\
\fl \left.+\ldots+\widetilde{C}_{\bm{n}_0}\frac{t^k}{k!}
\left(-K_{\mathrm{out}}\pi(\bar{\bm{n}})\right)^k+
+\ldots 
+\widetilde{C}_{\bm{n}_0}\frac{t^{\bar{k}_t}}{\bar{k}_t!}\left(-K_{\mathrm{out}}
\pi(\bar{\bm{n}})\right)^{\bar{k}_t}
\right)
+\mathcal{R}_t,
\end{eqnarray} 
and must coincide with the analytic continuation of Eq. (\ref{Q8}) for $\bar{V}\to\widetilde{E}$.
According to Eq. (\ref{E0}), we have now to analyze the behavior of the rhs of Eqs. (\ref{Q8}) and (\ref{Q9})
in the limit $t\to\infty$ (note that, if we want to calculate the GS,   
we are allowed to take the thermodynamic limit only after performing the limit $t\to\infty$). 
Were the series involved in these equations truncated at a finite
number of $k$, we could get immediately $|E\rangle$ and $E$, by simply choosing between 
$|\widetilde{E}\rangle$ and $\widetilde{E}$, or $|\bar{\bm{n}}\rangle$ and $\bar{V}$,
according to which is the smallest one between $\widetilde{E}$ or $\bar{V}$.
This would erroneously bring us to claim that the phase transition scenario summarized in Sec. II
occurs actually also for finite sizes! Such an error of course depends on the fact that
we have truncated the series. In fact, when we have to sum an infinite number of terms
of the form $\exp(\alpha_k t)$, in general, the resulting sum will be a new exponential in $t$ but
having a new suitable exponent.     
We can better understand this phenomena from the easier case $\bar{V}=\widetilde{E}$, \textit{i.e.},
near the critical point.
In this case we are able to sum the series in Eq. (\ref{Q9}) and we get 
$E=\bar{V}+\bar{\pi}K_{\mathrm{out}}$ and not $E=\bar{V}$.
We see therefore that: \textit{i)} only in the limit $\bar{\pi}\to 0$ we are allowed to claim from
Eqs. (\ref{Q8}) and (\ref{Q9}) which is the GS and the GS energy by simply taking
the smallest between $\widetilde{E}$ or $\bar{V}$, obtaining then the phase transition scenario
described in Sec. II; \textit{ii)} there is no phase separation
for $\bar{\pi}$ finite; \textit{iii)} at the critical point, or in a sufficiently small region
near the critical point, for small but finite $\bar{\pi}$, the GS  
is always a mixed phase to which contribute both the reservoir and the cavity,
and $E-\widetilde{E}\simeq E-\bar{V}$ is of the order $\mathop{O}(\bar{\pi}K_{\mathrm{out}})$, 
so that, given $\bar{\pi}$, the bigger is $K_{\mathrm{out}}$,
the stronger will be the finite size effects. 
Note finally that, due to Eq. (\ref{PIb}), all these statements can be reformulated in terms of
of $\bar{p}=1/|\mathbb{F}|$, being $\bar{p}N\to 0$ a sufficient condition for having $\bar{\pi}N\to 0$.

\subsection{Proof of Eqs. (\ref{THEO8})-(\ref{THEO10}) (exact derivation)}
We now generalize the above derivation to the case in which the cavity $\bar{\mathbb{F}}$
is any infinitesimal subset of the Fock space such that $\bar{\pi}N\to 0$ for $N\to\infty$. 
It is convenient to define the internal boundaries of the reservoir and cavity sets as
\begin{eqnarray}
\label{Q20} 
\partial\bar{\mathbb{F}}\defi \cup_{\bm{n}\in \bar{\mathbb{F}}: ~\exists \bm{n}'\notin\bar{\mathbb{F}}: ~K_{\bm{n},\bm{n}'}\neq 0}\bm{n},
\end{eqnarray}
\begin{eqnarray}
\label{Q20r} 
\partial\widetilde{\mathbb{F}}\defi 
\cup_{\bm{n}\in \widetilde{\mathbb{F}}: ~\exists \bm{n}'\notin\widetilde{\mathbb{F}}: ~K_{\bm{n},\bm{n}'}\neq 0}\bm{n}.
\end{eqnarray}
For each configuration $\bm{n}\in \bar{\mathbb{F}}$ we decompose 
the total number of active links $A(\bm{n})$ as 
\begin{eqnarray}
\label{Q21}
A(\bm{n})=A_{\mathrm{in}}(\bm{n})+A_{\mathrm{out}}(\bm{n}), 
\end{eqnarray}
where $A_{\mathrm{in}}(\bm{n})$ and $A_{\mathrm{out}}(\bm{n})$ represent the number
of connections of $\bm{n}$ to configurations belonging or not belonging to $\bar{\mathbb{F}}$, respectively.
More explicitly
\begin{eqnarray}
\label{Q22}
A_{\mathrm{in}}(\bm{n})&\defi& \sum_{\bm{n}'\in \bar{\mathbb{F}}}|\lambda_{\bm{n},\bm{n}'}|, \\
A_{\mathrm{out}}(\bm{n})&\defi& \sum_{\bm{n}'\notin \bar{\mathbb{F}}}|\lambda_{\bm{n},\bm{n}'}|.
\end{eqnarray}
Clearly, if $\bm{n}\in \bar{\mathbb{F}}\setminus\partial\bar{\mathbb{F}}$ then $A_{\mathrm{out}}(\bm{n})=0$.
Similarly, for $\bm{n}\in \bar{\mathbb{F}}$, we decompose the weighted number (or kinetic term) $R(\bm{n})$ as
\begin{eqnarray}
\label{Q23}
R(\bm{n})=R_{\mathrm{in}}(\bm{n})+R_{\mathrm{out}}(\bm{n}), 
\end{eqnarray}
where
\begin{eqnarray}
\label{Q24}
R_{\mathrm{in}}(\bm{n})&\defi& \sum_{\bm{n}'\in \bar{\mathbb{F}}}|K_{\bm{n},\bm{n}'}|, \\
\label{Q24b}
R_{\mathrm{out}}(\bm{n})&\defi& \sum_{\bm{n}'\notin \bar{\mathbb{F}}}|K_{\bm{n},\bm{n}'}|.
\end{eqnarray}
The corresponding definitions for the reservoir space are understood. Sometimes, to stress 
whether the kinetic terms $R_{\mathrm{in}}$ and $R_{\mathrm{out}}$ refer to the cavity, or to the reservoir,
we will write $\bar{R}_{\mathrm{in}}$ and $\bar{R}_{\mathrm{out}}$, or
$\widetilde{R}_{\mathrm{in}}$ and $\widetilde{R}_{\mathrm{out}}$, respectively. 
Note that the correct kinetic terms associated to the Hamiltonian's
$\bar{H}$ and $\widetilde{H}$ of the cavity and reservoir spaces, 
are $\bar{R}_{\mathrm{in}}(\bm{n})$ and $\widetilde{R}_{\mathrm{in}}(\bm{n})$,
respectively.
 
Let $t_1,\ldots,t_k$ be the stochastic times at which the system enters in the subspace $\bar{\mathbb{F}}$,
and let $\tau_1,\ldots,\tau_k$ be the corresponding living times.
Note that, by construction, at the times $t_1^+,\ldots,t_k^+$, as well as at the times 
$t_1+\tau_1^-,\ldots,t_k+\tau_k^-$, the system is on the boundary $\partial\bar{\mathbb{F}}$.
As in the previous paragraph we can repeat the same steps also for the present degenerate case,
Eqs. (\ref{Q1}) and (\ref{Q2}) being formally identical with the obvious generalizations
for the symbols $Q_t(k)$, $\widetilde{\mathcal{M}}^{[0,t)}$ and $\bar{\mathcal{M}}^{[0,t)}$.
Note however that $\widetilde{\mathcal{M}}^{[0,t)}$ and $\bar{\mathcal{M}}^{[0,t)}$, defined
as restrictions of the stochastic functional $\mathcal{M}^{[0,t)}$ to the configurations
of the spaces $\widetilde{\mathbb{F}}$ and $\bar{\mathbb{F}}$, respectively,
are not yet associated to the Hamiltonian's $\widetilde{H}$ and $\bar{H}$ defined as the
restriction of $H$ to these spaces.
In fact, the kinetic terms $R$'s appearing in these stochastic functionals
are different from the kinetic terms $R_{\mathrm{in}}$'s associated to $\widetilde{H}$ or $\bar{H}$
when the system is on the boundary of $\widetilde{\mathbb{F}}$  
($\partial\widetilde{\mathbb{F}}$) or
on the boundary of $\bar{\mathbb{F}}$ ($\partial\bar{\mathbb{F}}$), respectively. 
Furthermore here, to repeat the analogous derivation of the previous paragraph, we should calculate
exactly the living times in the subspace $\bar{\mathbb{F}}$ with respect to which we have
explicited the expectation (\textit{i.e.}, the analogous of Eq. (\ref{Q3})).
Let us first consider the simplest case in which for any $\bm{n}\in\bar{\mathbb{F}}$ we have 
a fixed value for both the potential, $V(\bm{n})\equiv\bar{V}$, 
and for $R(\bm{n})\equiv\bar{R}$, $R_{\mathrm{in}}(\bm{n})\equiv\bar{R}_{\mathrm{in}}$, 
and then also $R_{\mathrm{out}}(\bm{n})\equiv\bar{R}_{\mathrm{out}}$.  
We observe now that, once at the time $\tau=0$ the system enters in a configuration $\bm{n}\in\bar{\mathbb{F}}$,
the infinitesimal probability that the system leaves $\bar{\mathbb{F}}$ after a time between $\tau$ and $\tau+d\tau$
is given by $d\tau \bar{R}_{\mathrm{out}}\exp(-\bar{R}_{\mathrm{out}}\tau)$. 
Therefore, according to Eqs. (\ref{EPRGC}) and  (\ref{Q23}) we see that the analogous of 
Eq. (\ref{Q3}) is given by (for simplicity in this example we consider a bosonic case so that the $\lambda$'s associated
to the jumping links are all positive)
\begin{eqnarray}
\label{Q25}\fl
\E\left(\mathcal{M}^{[0,t)}\right)&=&\sum_{k=0}^{\infty}\int_{0}^{'t} dx_1dx_2\ldots dx_k\E\left(
\widetilde{\mathcal{M}}^{[0,t_1)}\widetilde{\mathcal{M}}^{[t_1+x_1,t_2)}\cdots 
\widetilde{\mathcal{M}}^{[t_k+x_k,t)} \right. \nonumber \\
&& \times \left.
e^{\left[-\bar{R}_{\mathrm{in}}+\bar{V}\right](x_1+x_2\ldots+x_k)}
|k_t=k; \tau_1=x_1,\ldots, \tau_k=x_k\right) \nonumber \\
&& \times  \bar{R}_{\mathrm{out}}^k Q_t(k).
\end{eqnarray}
We see then that, similarly to
the non degenerate case, the probabilities of the living times in $\bar{\mathbb{F}}$ cancel
exactly with the portions of the kinetic terms $\bar{R}_{\mathrm{out}}$, while the rest of the kinetic terms, \textit{i.e.} 
$\bar{R}_{\mathrm{in}}$,
are exactly those associated to the true $\bar{H}$ (note that in this example $\bar{E}=-\bar{R}_{\mathrm{in}}+\bar{V}$).
Now, for the general case, analogously to the previous paragraph, 
we can write (we suppose to start from a configuration in $\widetilde{\mathbb{F}}$)
\begin{eqnarray}
\label{Q3bg}\fl
\E\left(\mathcal{M}^{[0,t)}\right)&=&\sum_{k=0}^{\infty}\E\left(
\widetilde{\mathcal{M}}_{\widetilde{\bm{n}}_{0}}^{[0,t_1)}\lambda_{\bm{n}_{t_1^-},\bm{n}_{t_1^+}} 
\bar{\mathcal{M}}_{\bar{\bm{n}}_{t_1^+}}^{[t_1,t_1+\tau_1)}\lambda_{\bm{n}_{t_1+\tau_1^-},\bm{n}_{t_1+\tau_1^+}}
\cdots \right.
\nonumber \\ && \times 
\left. \bar{\mathcal{M}}_{\bar{\bm{n}}_{t_k^+}}^{[t_k,t_k+\tau_k)}
\lambda_{\bm{n}_{t_k+\tau_k^-},\bm{n}_{t_k+\tau_k^+}}
\widetilde{\mathcal{M}}_{\widetilde{\bm{n}}_{t_k+\tau_k^+}}^{[t_k+\tau_k,t)}
|k_t=k \right)Q_t(k),
\end{eqnarray}
We would like now to generalize Eq. (\ref{Q25}).
However, the probability of the living times in $\bar{\mathbb{F}}$ for the general case represents
a formidable task. Nevertheless, as we shall see, for our aim we do not need to calculate
this quantity, but only a suitable conditional expectation.
In fact we can use the following Lemma.

\textit{Lemma.}\\
Let $\tau$ be a generic random time at which the system, for the first time before $t$, leaves some
region of the Fock space
(for example $\tau$ can be the first time at which the system leaves, or enters, the cavity). 
Then we have
\begin{eqnarray}
\label{QLemma1}
\E\left(\mathcal{M}^{[0,\tau)}_{\bm{n}_0}\right)&=&\int_0^t dx~\E\left(\mathcal{M}^{[0,x)}_{\mathrm{in};\bm{n}_0}
R_{\mathrm{out}}(\bm{n_{x^-}})|\tau=x\right),
\end{eqnarray}
where the symbol $\mathcal{M}^{[0,x)}_{\mathrm{in};\bm{n}_0}$ indicates that 
in the stochastic functional now the kinetic terms are only those internal to the region:    
\begin{eqnarray}
\label{QDEF}
\mathcal{M}^{[0,t)}_{\mathrm{in};\bm{n}_0}\defi
e^{\int_0^t\left[R_{\mathrm{in}}\left(\bm{n}_s\right)-V\left(\bm{n}_s\right)\right]ds} 
\prod_{k=1}^{N_{t}} \lambda_k.
\end{eqnarray}
More in general, for what we need, we have
\begin{eqnarray}
\label{QLemma2}\fl
\E\left(\widetilde{\mathcal{M}}_{\widetilde{\bm{n}}_{0}}^{[0,t_1)}\lambda_{\bm{n}_{t_1^-},\bm{n}_{t_1^+}} 
\bar{\mathcal{M}}_{\bar{\bm{n}}_{t_1^+}}^{[t_1,t_1+\tau_1)}\lambda_{\bm{n}_{t_1+\tau_1^-},\bm{n}_{t_1+\tau_1^+}} \cdots \right.
\nonumber \\ \fl \times \left. 
\bar{\mathcal{M}}_{\bar{\bm{n}}_{t_k^+}}^{[t_k,t_k+\tau_k)}
\lambda_{\bm{n}_{t_k+\tau_k^-},\bm{n}_{t_k+\tau_k^+}}
\widetilde{\mathcal{M}}_{\widetilde{\bm{n}}_{t_k+\tau_k^+}}^{[t_k+\tau_k,t)}
|k_t=k \right)= \nonumber \\
\fl \int_0^{'t} dx_1\ldots dx_k~\E\left(
\widetilde{\mathcal{M}}_{\mathrm{in};\widetilde{\bm{n}}_{0}}^{[0,t_1)}\lambda_{\bm{n}_{t_1^-},\bm{n}_{t_1^+}} 
\bar{\mathcal{M}}_{\mathrm{in};\bar{\bm{n}}_{t_1^+}}^{[t_1,t_1+x_1)}R_{\mathrm{out}}(\bm{n}_{t_1+x_1^-})\lambda_{\bm{n}_{t_1+x_1^-},\bm{n}_{t_1+\tau_1^+}}
\cdots \right.
\nonumber \\ \fl \times
\left. \bar{\mathcal{M}}_{\mathrm{in};\bar{\bm{n}}_{t_k}^+}^{[t_k,t_k+x_k)}
R_{\mathrm{out}}(\bm{n}_{t_k+x_k^-})\lambda_{\bm{n}_{t_k+x_k^-},\bm{n}_{t_k+x_k^+}} \right.
\nonumber \\ \fl \times 
\left. \widetilde{\mathcal{M}}_{\mathrm{in};\widetilde{\bm{n}}_{t_k+x_k^+}}^{[t_k+x_k,t)}
|k_t=k;\tau_1=x_1,\ldots,\tau_k=x_k \right).
\end{eqnarray}

Furthermore we have the following ergodic property
\begin{eqnarray}
\label{Qergo1}
\lim_{x_1\to\infty}\frac{\E\left(\bar{\mathcal{M}}^{[0,x_1)}_{\mathrm{in};\bar{\bm{n}}_0}|\tau_1=x_1\right)}
{\bar{\E}\left(\bar{\mathcal{M}}^{[0,x_1)}_{\mathrm{in};\bar{\bm{n}}_0}\right)}=1,
\end{eqnarray}
and similarly
\begin{eqnarray}
\label{Qergo2}
\lim_{y_1\to\infty}\frac{\E\left(\widetilde{\mathcal{M}}^{[0,y_1)}_{\mathrm{in};\widetilde{\bm{n}}_0}|t_1=y_1\right)}
{\E\left(\widetilde{\mathcal{M}}^{[0,y_1)}_{\mathrm{in};\widetilde{\bm{n}}_0}\right)}=1.
\end{eqnarray}

The proof of this Lemma is reported in Appendix.

Let us now exploit the Lemma for the case $k_t=1$ \cite{NOTE6}.
By using the Markov property we can rewrite this contribution as \cite{SHY} 
\begin{eqnarray}
\label{QLemma3}\fl
\E\left(
\widetilde{\mathcal{M}}_{\mathrm{in};\widetilde{\bm{n}}_{0}}^{[0,t_1)}\lambda_{\bm{n}_{t_1^-},\bm{n}_{t_1^+}} 
\bar{\mathcal{M}}_{\mathrm{in};\bar{\bm{n}}_{t_1^+}}^{[t_1,t_1+x_1)}R_{\mathrm{out}}(\bm{n}_{t_1+x_1^-})\lambda_{\bm{n}_{t_1+x_1^-},\bm{n}_{t_1+\tau_1^+}} \right.
\nonumber \\ \fl \left. \times \widetilde{\mathcal{M}}_{\mathrm{in};\widetilde{\bm{n}}_{t_1+x_1^+}}^{[t_1+x_1,t)}
|k_t=1;\tau_1=x_1\right)=\\
\fl \E\left(
\widetilde{\mathcal{M}}_{\mathrm{in};\widetilde{\bm{n}}_{0}}^{[0,t_1)}\lambda_{\bm{n}_{t_1^-},\bm{n}_{t_1^+}} 
\bar{\mathcal{M}}_{\mathrm{in};\bar{\bm{n}}_{t_1^+}}^{[t_1,t_1+x_1)}R_{\mathrm{out}}(\bm{n}_{t_1+x_1^-})\lambda_{\bm{n}_{t_1+x_1^-},\bm{n}_{t_1+\tau_1^+}} \right.
\nonumber \\ \fl \left. \times 
\E\left(\widetilde{\mathcal{M}}_{\mathrm{in};\widetilde{\bm{n}}_{t_1+x_1^+}}^{[t_1+x_1,t)}|\widetilde{\bm{n}}_{t_1+x_1}\right)
|k_t=1;\tau_1=x_1\right).
\end{eqnarray} 
The last conditional expectation that appears in the rhs of the above expression is a random
variable that, using the EPR, can be written as
\begin{eqnarray}
\label{QLemma4}\fl
\E\left(\widetilde{\mathcal{M}}_{\mathrm{in};\widetilde{\bm{n}}_{t_1+x_1^+}}^{[t_1+x_1,t)}|\widetilde{\bm{n}}_{t_1+x_1^+}\right)=
\widetilde{C}_{\widetilde{\bm{n}}_{t_1+x_1^+}}e^{-\widetilde{E}[t-(t_1+x_1)]}+\mathop{O}\left(e^{-\widetilde{E}_1[t-(t_1+x_1)]}\right),
\end{eqnarray} 
where $\widetilde{E}_1$ is the first excited state of the reservoir, and
$\widetilde{C}_{\widetilde{\bm{n}}}$ is given by
\begin{eqnarray}
\label{QLemma5}
\widetilde{C}_{\widetilde{\bm{n}}}\defi \sum_{\bm{n}'}\langle \bm{n}'|\widetilde{E}\rangle 
\langle \widetilde{E} |\widetilde{\bm{n}}\rangle.
\end{eqnarray} 
Note that, since $|\widetilde{E}\rangle$ is normalized to 1,
it is $\widetilde{C}_{\widetilde{\bm{n}}}=\mathop{O}\left(1\right)$.
By inserting Eq. (\ref{QLemma4}) in Eq. (\ref{QLemma3}), up to terms exponentially smaller 
in $t-(t_1+x_1)$, 
and by using again the Markov property, we get
\begin{eqnarray}
\label{QLemma6}
\fl \E\left(
\widetilde{\mathcal{M}}_{\mathrm{in};\widetilde{\bm{n}}_{0}}^{[0,t_1)}\lambda_{\bm{n}_{t_1^-},\bm{n}_{t_1^+}} 
\bar{\mathcal{M}}_{\mathrm{in};\bar{\bm{n}}_{t_1^+}}^{[t_1,t_1+x_1)}R_{\mathrm{out}}(\bm{n}_{t_1+x_1^-})\lambda_{\bm{n}_{t_1+x_1^-},\bm{n}_{t_1+\tau_1^+}} \right.
\nonumber \\ \fl \left. \times 
\E\left(\widetilde{\mathcal{M}}_{\mathrm{in};\widetilde{\bm{n}}_{t_1+x_1^+}}^{[t_1+x_1,t)}|\widetilde{\bm{n}}_{t_1+x_1}\right)
|k_t=1;\tau_1=x_1\right)=\nonumber\\
\fl
\E\left(e^{-\widetilde{E}[t-(t_1+x_1)]}
\widetilde{\mathcal{M}}_{\mathrm{in};\widetilde{\bm{n}}_{0}}^{[0,t_1)}\lambda_{\bm{n}_{t_1^-},\bm{n}_{t_1^+}} \right.
\times \nonumber \\ \fl \left. 
\E\left(\bar{\mathcal{M}}_{\mathrm{in};\bar{\bm{n}}_{t_1^+}}^{[t_1,t_1+x_1)}R_{\mathrm{out}}(\bm{n}_{t_1+x_1^-})\lambda_{\bm{n}_{t_1+x_1^-},\bm{n}_{t_1+\tau_1^+}} 
\widetilde{C}_{\widetilde{\bm{n}}_{t_1+x_1^+}}|\bm{n}_{t_1^+}\right)
|k_t=1;\tau_1=x_1\right).
\end{eqnarray} 
In turn, the last expression can be expressed in terms of the probability that
given $\bm{n}_{t_1+x_1^-}$, the system goes from $\bm{n}_{t_1+x_1^-}$ to a generic configuration 
$\bm{n}\in\widetilde{\mathbb{F}}$. 
Taking into account Eqs. (\ref{Kf})-(\ref{Af}), we have
\begin{eqnarray}
\label{QLemma7} \fl
\E\left(\bar{\mathcal{M}}_{\mathrm{in};\bar{\bm{n}}_{t_1^+}}^{[t_1,t_1+x_1)}R_{\mathrm{out}}(\bm{n}_{t_1+x_1^-})\lambda_{\bm{n}_{t_1+x_1^-},\bm{n}_{t_1+\tau_1^+}} 
\widetilde{C}_{\widetilde{\bm{n}}_{t_1+x_1^+}}|\bm{n}_{t_1^+};\tau_1=x_1\right)=\nonumber \\
\fl \E\left(\bar{\mathcal{M}}_{\mathrm{in};\bar{\bm{n}}_{t_1^+}}^{[t_1,t_1+x_1)}
\sum_{\bm{n}}\frac{-R_{\mathrm{out}}(\bm{n}_{t_1+x_1^-})K_{\bm{n}_{t_1+x_1^-},\bm{n}}}
{R_{\mathrm{out}}(\bm{n}_{t_1+x_1^-})} ~\widetilde{C}_{\bm{n}}|\bm{n}_{t_1^+};\tau_1=x_1\right)=
\nonumber \\ \fl 
\E\left(\bar{\mathcal{M}}_{\mathrm{in};\bar{\bm{n}}_{t_1^+}}^{[t_1,t_1+x_1)}|\bm{n}_{t_1^+};\tau_1=x_1\right)
\langle\sum_{\bm{n}}-K_{\bm{n}_{t_1+x_1^-},\bm{n}} ~\widetilde{C}_{\bm{n}}\rangle_{\mathrm{cavity}},
\end{eqnarray} 
where, in the last expression, we have introduced the normalized average
\begin{eqnarray}
\label{QLemma8} \fl
\langle\sum_{\bm{n}}K_{\bm{n}_{t_1+x_1^-},\bm{n}}~\widetilde{C}_{\bm{n}}\rangle_{\mathrm{cavity}}\defi
\nonumber \\ \fl 
\frac{\E\left(\bar{\mathcal{M}}_{\mathrm{in};\bar{\bm{n}}_{t_1^+}}^{[t_1,t_1+x_1)}
\sum_{\bm{n}}K_{\bm{n}_{t_1+x_1^-},\bm{n}}~\widetilde{C}_{\bm{n}}|\bm{n}_{t_1^+};\tau_1=x_1\right)}
{\E\left(\bar{\mathcal{M}}_{\mathrm{in};\bar{\bm{n}}_{t_1^+}}^{[t_1,t_1+x_1)}|\bm{n}_{t_1^+};\tau_1=x_1\right)}.
\end{eqnarray} 
Note, in Eq. (\ref{QLemma7}), the presence of the denominator $\bar{R}_{\mathrm{out}}$.
This is due to the fact that, the probability to be used in Eq. (\ref{QLemma7})
must take into account that the expectation to be calculated is conditioned by 
$\tau_1=x_1$, \textit{i.e.}, the system at the time $x_1$ leaves the cavity with probability 1, 
and, with a jump, can go to any of the $R_{\mathrm{out}}(\bm{n}_{t_1+x_1^-})$ configurations of the reservoir space
(in the non degenerate case $\bar{R}_\mathrm{out}=\bar{R}$).

Note that, as $x_1$ goes to $\infty$, the normalized average (\ref{QLemma8}) 
stabilizes exponentially fast to a asymptotic value 
independent on the initial condition $\bm{n}_{t_1^+}$.
Therefore, taking into account that eventually we need only to keep track 
of the behavior for $t\to\infty$, and then also $x_1\to\infty$,
up to exponentially smaller terms, the rhs of Eq. (\ref{QLemma6}) becomes now
\begin{eqnarray}
\label{QLemma9}
\fl
\E\left(e^{-\widetilde{E}[t-(t_1+x_1)]}
\widetilde{\mathcal{M}}_{\mathrm{in};\widetilde{\bm{n}}_{0}}^{[0,t_1)}\lambda_{\bm{n}_{t_1^-},\bm{n}_{t_1^+}} 
\E\left(\bar{\mathcal{M}}_{\mathrm{in};\bar{\bm{n}}_{t_1^+}}^{[t_1,t_1+x_1)}|\bm{n}_{t_1^+};\tau_1=x_1\right)
|k_t=1;\tau_1=x_1\right)
\nonumber \\ \fl \times
\lim_{t\to\infty}\langle\sum_{\bm{n}}-K_{\bm{n}_{t^-},\bm{n}} 
\widetilde{C}_{\bm{n}}\rangle_{\mathrm{cavity}}.
\end{eqnarray} 
Similarly to what we have done before in Eq. (\ref{QLemma4}), 
we see that the last conditional expectation that appears in the above expression is a random
variable that, using again the EPR and Eq. (\ref{Qergo1}), can be written as
\begin{eqnarray}
\label{QLemma10}\fl
\E\left(\bar{\mathcal{M}}_{\mathrm{in};\bar{\bm{n}}_{t_1^+}}^{[t_1,t_1+x_1)}|\bar{\bm{n}}_{t_1^+};\tau_1=x_1\right)=
\bar{C}_{\bar{\bm{n}}_{t_1^+}}e^{-\bar{E}x_1}+\mathop{O}\left(e^{-\bar{E}_1x_1}\right),
\end{eqnarray} 
where $\bar{E}_1$ is the first excited state of the cavity, and
$\bar{C}_{\bar{\bm{n}}}$ is given by
\begin{eqnarray}
\label{QLemma11}
\bar{C}_{\bar{\bm{n}}}\defi \sum_{\bm{n}'}\langle \bm{n}'|\bar{E}\rangle 
\langle \bar{E} |\bar{\bm{n}}\rangle.
\end{eqnarray} 
Note that, since $|\bar{E}\rangle$ is normalized to 1,
it is $\bar{C}_{\bar{\bm{n}}}=\mathop{O}\left(1\right)$.
By inserting Eq. (\ref{QLemma10}) in Eq. (\ref{QLemma9}) we obtain
\begin{eqnarray}
\label{QLemma12}
\fl
\E\left(e^{-\widetilde{E}[t-(t_1+x_1)]}
\widetilde{\mathcal{M}}_{\mathrm{in};\widetilde{\bm{n}}_{0}}^{[0,t_1)}\lambda_{\bm{n}_{t_1^-},\bm{n}_{t_1^+}} 
\E\left(\bar{\mathcal{M}}_{\mathrm{in};\bar{\bm{n}}_{t_1^+}}^{[t_1,t_1+x_1)}|\bm{n}_{t_1^+};\tau_1=x_1\right)
|k_t=1;\tau_1=x_1\right)=\nonumber \\
\fl \E\left(e^{-\widetilde{E}[t-(t_1+x_1)]-\bar{E}x_1}
\widetilde{\mathcal{M}}_{\mathrm{in};\widetilde{\bm{n}}_{0}}^{[0,t_1)}\lambda_{\bm{n}_{t_1^-},\bm{n}_{t_1^+}} 
\bar{C}_{\bar{\bm{n}}_{t_1^+}}
|k_t=1;\tau_1=x_1\right).
\end{eqnarray} 
Finally, by performing analogous steps as those done in
Eqs. (\ref{QLemma7})-(\ref{QLemma9}), and by using Eq. (\ref{Qergo2}), we arrive at
\begin{eqnarray}
\label{QLemma13}
\fl \E\left(e^{-\widetilde{E}[t-(t_1+x_1)]-\bar{E}x_1}
\widetilde{\mathcal{M}}_{\mathrm{in};\widetilde{\bm{n}}_{0}}^{[0,t_1)}\lambda_{\bm{n}_{t_1^-},\bm{n}_{t_1^+}} 
\bar{C}_{\bar{\bm{n}}_{t_1^+}}
|k_t=1;\tau_1=x_1\right)=\widetilde{C}_{\bm{n}_0}\times \nonumber \\
\fl \E\left(e^{-\widetilde{E}[t-(t_1+x_1)]-\bar{E}x_1-\widetilde{E}t_1}|k_t=1;\tau_1=x_1\right)
\lim_{t\to\infty}\langle\sum_{\bm{n}}\frac{-K_{\bm{n}_{t^-},\bm{n}}}
{\widetilde{R}_{\mathrm{out}}(\bm{n}_{t^-})} 
\bar{C}_{\bm{n}}\rangle_{\mathrm{reservoir}},
\end{eqnarray} 
where we have introduced the normalized average (see note \cite{NOTE6})
\begin{eqnarray}
\label{QLemma14} \fl
\langle\sum_{\bm{n}}\frac{-K_{\bm{n}_{y_1^-},\bm{n}}}{\widetilde{R}_{\mathrm{out}}(\bm{n}_{y_1^-})}  
~\bar{C}_{\bm{n}}\rangle_{\mathrm{reservoir}}\defi
\frac{\E\left(\widetilde{\mathcal{M}}_{\mathrm{in};\widetilde{\bm{n}}_{0}}^{[0,y_1)}
\sum_{\bm{n}} \frac{-K_{\bm{n}_{y_1^-},\bm{n}}}{\widetilde{R}_{\mathrm{out}}(\bm{n}_{y_1^-})}~\bar{C}_{\bm{n}}|
\bm{n}_{0};k_t=1;t_1=y_1\right)}
{\E\left(\bar{\mathcal{M}}_{\mathrm{in};\bar{\bm{n}}_{0}}^{[0,y_1)}|\bm{n}_{0};k_t=1;t_1=y_1\right)}.
\end{eqnarray} 
Note, in Eq. (\ref{QLemma14}), the presence of the denominator $\widetilde{R}_{\mathrm{out}}$.
This is due to the fact that, the probability to be used in Eq. (\ref{QLemma13}),
must take into account that the expectation to be calculated is conditioned by 
$t_1=y_1$, \textit{i.e.}, the system at the time $y_1$ leaves the reservoir with probability 1, 
and, with a jump, can go to any of the $\widetilde{R}_{\mathrm{out}}(\bm{n}_{t_1^-})$ configurations of the cavity space
(in the non degenerate case $\widetilde{R}_{\mathrm{out}}(\bm{n}_{t_1^-})\equiv 1$).

Putting all together from Eq. (\ref{QLemma3}) to Eq. (\ref{QLemma13}) we hence obtain
\begin{eqnarray}
\label{QLemma14b}\fl
\E\left(
\widetilde{\mathcal{M}}_{\mathrm{in};\widetilde{\bm{n}}_{0}}^{[0,t_1)}\lambda_{\bm{n}_{t_1^-},\bm{n}_{t_1^+}} 
\bar{\mathcal{M}}_{\mathrm{in};\bar{\bm{n}}_{t_1^+}}^{[t_1,t_1+x_1)}R_{\mathrm{out}}(\bm{n}_{t_1+x_1^-})\lambda_{\bm{n}_{t_1+x_1^-},\bm{n}_{t_1+\tau_1^+}} \right.
\nonumber \\ \fl \left. \times \widetilde{\mathcal{M}}_{\mathrm{in};\widetilde{\bm{n}}_{t_1+x_1^+}}^{[t_1+x_1,t)}
|k_t=1;\tau_1=x_1\right)=\widetilde{C}_{\bm{n}_0}e^{-\widetilde{E}t+(\widetilde{E}-\bar{E})x_1}
\\ \fl \times
\lim_{t\to\infty}\langle\sum_{\bm{n}}K_{\bm{n}_{t^-},\bm{n}}
\widetilde{C}_{\bm{n}}\rangle_{\mathrm{cavity}}
\lim_{t\to\infty}\langle\sum_{\bm{n}}\frac{K_{\bm{n}_{t^-},\bm{n}}}
{\widetilde{R}_{\mathrm{out}}(\bm{n}_{t^-})} 
\bar{C}_{\bm{n}}\rangle_{\mathrm{reservoir}}.
\end{eqnarray} 
Notice, at this leading order, the null role played by the random variable $t_1$. 

In general, by iterating $k$ times the above procedure,
up to terms exponentially smaller in the random variables $t$ and $(x_1+\ldots +x_k)$, we get 
\begin{eqnarray}
\label{QLemma15}
\fl \E\left(
\widetilde{\mathcal{M}}_{\mathrm{in};\widetilde{\bm{n}}_{0}}^{[0,t_1)}\lambda_{\bm{n}_{t_1^-},\bm{n}_{t_1^+}} 
\bar{\mathcal{M}}_{\mathrm{in};\bar{\bm{n}}_{t_1^+}}^{[t_1,t_1+x_1)}R_{\mathrm{out}}(\bm{n}_{t_1+x_1^-})\lambda_{\bm{n}_{t_1+x_1^-},\bm{n}_{t_1+\tau_1^+}}
\cdots \bar{\mathcal{M}}_{\mathrm{in};\bar{\bm{n}}_{t_k}^+}^{[t_k,t_k+x_k)}\right.
\nonumber \\ \fl \times
\left. 
R_{\mathrm{out}}(\bm{n}_{t_k+x_k^-})\lambda_{\bm{n}_{t_k+x_k^-},\bm{n}_{t_k+x_k^+}}
\widetilde{\mathcal{M}}_{\mathrm{in};\widetilde{\bm{n}}_{t_k+x_k^+}}^{[t_k+x_k,t)}
|k_t=k;\tau_1=x_1,\ldots,\tau_k=x_k \right)= \nonumber \\ 
\fl \widetilde{C}_{\bm{n}_0}e^{-\widetilde{E}t+(\widetilde{E}-\bar{E})(x_1+\ldots+x_k)}(-K_{\mathrm{out}})^k,
\end{eqnarray}
where 
\begin{eqnarray}
\label{QLemma16}\fl
K_{\mathrm{out}}\defi 
-\lim_{t\to\infty}\langle\sum_{\bm{n}}K_{\bm{n}_{t^-},\bm{n}}
\widetilde{C}_{\bm{n}}\rangle_{\mathrm{cavity}}
\lim_{t\to\infty}\langle\sum_{\bm{n}}\frac{K_{\bm{n}_{t^-},\bm{n}}}
{\widetilde{R}_{\mathrm{out}}(\bm{n}_{t^-})} 
\bar{C}_{\bm{n}}\rangle_{\mathrm{reservoir}}.
\end{eqnarray}

We can now insert Eqs. (\ref{QLemma15}) and (\ref{QLemma16}) in Eq. (\ref{QLemma2}).
Finally, from Eq. (\ref{Q3bg}), up to terms exponentially smaller in $t$, 
we arrive at the analogous of Eq. (\ref{Q5}) given now by
\begin{eqnarray}
\label{QLemma17}\fl
\E\left(\mathcal{M}^{[0,t)}\right)&=& \widetilde{C}_{\bm{n}_0} e^{-\widetilde{E}t}
\sum_{k=0}^{\infty}\int_{0}^{'t} dx_1dx_2\ldots dx_k
e^{(\widetilde{E}-\bar{E})(x_1+\ldots+x_k)} \left(-K_{\mathrm{out}}\right)^k Q_t(k).
\end{eqnarray}
We see that Eq. (\ref{QLemma17}) is just the wanted generalization of the specific case 
we had analyzed before
in which for any $\bm{n}\in\bar{\mathbb{F}}$ we had 
a fixed number for $R(\bm{n})=\bar{R}$, $R_{\mathrm{in}}(\bm{n})=\bar{R}_{\mathrm{in}}$, 
and $R_{\mathrm{out}}(\bm{n})=\bar{R}_{\mathrm{out}}$ (see Eq. (\ref{Q25})).

The phase transition scenario described in the theorem by 
Eqs. (\ref{THEO8})-(\ref{THEO10}) follows with the same identical argument given in the
previous paragraph (the non degenerate case).
We see in particular that, for a finite system and then a finite value 
$\bar{\pi}$, 
near the critical point, the difference between the
energy of the full system $E$, and the energy of the reduced systems $\bar{E}$ or $\widetilde{E}$, is of the order
$\mathop{O}(K_{\mathrm{out}}\bar{\pi})$.
We observe that, using the fact that the GS's $|\widetilde{E}\rangle$ and $|\bar{E}\rangle$
are normalized to 1, from Eq. (\ref{QLemma16}), we have approximately 
\begin{eqnarray}
\label{QLemma18}
K_{\mathrm{out}}\sim -\langle\bar{R}_{\mathrm{out}}^+-\bar{R}_{\mathrm{out}}^-\rangle_{\mathrm{cavity}},
\end{eqnarray}
where $\bar{R}_{\mathrm{out}}^+$ and $\bar{R}_{\mathrm{out}}^-$, are the kinetic terms
obtained as weighted sums over the positive and negative active out-links, respectively.
Since, in general, $-(\langle{R}^+-{R}^-\rangle)$
is just the kinetic energy of the system \cite{BPDAJL}, 
we see that the meaning of $K_{\mathrm{out}}$ is that of the kinetic energy
by which the system, once entered in the cavity with an asymptotic probability $\bar{\pi}$,
is expelled from the cavity itself.
Given a small but finite value $\bar{\pi}$, 
the lower is this energy, the greater will be the ability of the system to remain
in the cavity for a longer time, making then faster (as $N$ grows) the approach to the phase transition.
Note also that, according to their definitions, the constants $\widetilde{C}_{\bm{n}}$ and
$\bar{C}_{\bm{n}}$, given in Eqs. (\ref{QLemma5}) and (\ref{QLemma11}), 
measure the overlap of the GS's $|\widetilde{E}\rangle$ and $|\bar{E}\rangle$ with the reservoir and cavity
internal boundaries $\partial\widetilde{\mathbb{F}}$ and $\partial\bar{\mathbb{F}}$, respectively.
Since, roughly speaking, these two overlaps appear as a product in $K_{\mathrm{out}}$,
we see that, as is intuitively evident, the larger are these overlaps, the stronger
will be the finite size effects and, more precisely, given the sum of the two overlaps, 
their effect will be maximal when
both the overlaps are relatively large, while minimal when one of them is relatively small.  

\section{An exact solvable example: The Random Potential Model}
In \cite{OP5} we have solved explicitly a model, called the Random Potential Model, a quantum quenched-disordered 
model, in which 
the potential levels $V_l$, with $l=1,\ldots,m$, take independent random values distributed according
to an arbitrary assigned distribution $p_l$, whereas the kinetic matrix $K$ is arbitrary
with the only constraint that its associated Markov chain be ergodic.
We have found that, up to small corrections that become negligible in the thermodynamic limit, the GS energy $E$ 
of this model satisfies:
\begin{eqnarray}
\label{E1f} 
\sum_{l}
 \frac{p_l}{E-V_l}=\frac{1}{E^{(0)}},  
\quad E\leq V_{1},
\end{eqnarray}
where $E^{(0)}$ is the GS energy of the system with zero potential ($V\equiv 0$),
and the constraint $E\leq V_{1}$ makes the solution unique.
We have then analyzed in detail this equation in the special limit $p_1\to 0$ and we have seen
that the system undergoes a QPT first-order phase transition in correspondence
of a suitable choice of the Hamiltonian parameters (\textit{i.e.}, the values of the $V_l$, their weights $p_l$
and the parameters entering the matrix elements of $K$ that affect then $E^{(0)}$).
Let us consider the simplest non trivial example with only two potential values $V_1$ and $V_2$.
In this case, Eq. (\ref{E1f}) is a quadratic equation for $e=E/N$
which can be solved explicitly.
Observing that $p_2=1-p_1$, and defining also $e^{(0)}=E^{(0)}/N$, $v_1=V_1/N$, and $v_2=V_2/N$, 
and taking into account that $e^{(0)}$ is negative,
one has \cite{OP5}
\begin{eqnarray}
\label{thermo2}\fl
e_0= v_1 - \case{1}{2}
\left[
\sqrt{(v_2-v_1+e^{(0)})^2 + 4 p_1 (v_2-v_1)}
-(v_2-v_1+e^{(0)})
\right],
\end{eqnarray}
the other solution being incompatible with the condition $e_0\leq v_1$.
We see that, for any $p_1>0$, $e_0$ is an analytic function of $v_2-v_1$ and $e^{(0)}$. 
However, for $p_1\to 0$ we have 
\begin{eqnarray}
\label{thermoe}\fl
\lim_{p_1 \to 0}e_0-v_1 &=  
\case{1}{2} \left[ (v_2-v_1+e^{(0)}) - |v_2-v_1+e^{(0)}| \right]
\nonumber \\
&= \cases{
v_2-v_1+e^{(0)}, & $v_2-v_1 < -e^{(0)}$, \\
0,   & $v_2-v_1 > -e^{(0)}$, \\
}
\end{eqnarray}
\textit{i.e.}, in the space of the Hamiltonian parameters 
$v_2-v_1$ and $e^{(0)}$, a singularity shows up at the critical point
\begin{eqnarray}
\label{critic}
v_2-v_1=-e^{(0)}. 
\end{eqnarray}
On the other hand, 
we see that Eq. (\ref{critic}) can be written also as
\begin{eqnarray}
\label{critic1}
\widetilde{e}=v_1. 
\end{eqnarray}
More in general in \cite{OP5}, we have shown that, for $p_1\to 0$, the critical condition
can be written as
\begin{eqnarray}
\label{COND}
W=\frac{1}{e^{(0)}},
\end{eqnarray}
where the function $W$ is defined as
\begin{eqnarray}
\label{thermoW}
W(p_2,v_2-v_1,\ldots,p_{m},v_{m}-v_1) \defi
\sum_{l=2}^{m} \frac{p_l}{v_1-v_l},
\end{eqnarray}
and $v_l\defi V_l/N$. 
Again, from Eqs. (\ref{E1f}) and (\ref{thermoW}), and taking into account that,
by definition, $p_1=0$ in the reservoir space, we see that the critical condition (\ref{COND}) 
is nothing else but the equality of the density energies $\widetilde{e}$ and $v_1$
that we have defined in our theorem for the finitely degenerate case.

An important application of this model is the Quantum Rem model, \textit{i.e.}, the Derrida's model 
\cite{PSPIN} immersed
in a transverse field, and the QPT here signals the transition of the system
between a quantum paramagnetic state and a glassy one \cite{BM}-\cite{Obuchi}.
These kind of models have important connections to the general problem of the
Quantum Annealing \cite{Krzakala}. 

For a general model, having then non random potential levels,
Eq. (\ref{E1f}) is of course not valid. However, 
it is possible to develop a perturbative approach whose
zero level approximation coincides with Eq. (\ref{E1f}), while non trivial
correlations of the potentials and the kinetic terms are taken into account
at higher order of the theory \cite{OP6}. Since the general structure of
the equations of this perturbative theory remains similar to Eq. (\ref{E1f}) itself,
we believe that a phase transition scenario similar to the one above described
will take place in the infinite dilution limit of suitable regions of the Fock space.

\section{A comment on the first exit times and the EPR}
Concerning the EPR and the random times $\tau$'s, as times at which the system leaves the cavity for the first
time before $t$, a comment is in order. From the EPR we know that to calculate the GS energy
of the system, we need to know the behavior of the expectation of the stochastic functional involved, only
in the limit $t\to\infty$. Nevertheless, in proving our theorem exactly, we needed a perfect
balance between weights and corresponding probabilities at each finite time interval. 
For example, from the formulae present in Appendix, it is not difficult to recognize
that, when $t\to\infty$, 
the pdf $p(\tau)$ for the random variable $\tau$ to leave
the cavity for the first time before $t$, satisfies the following asymptotic law
\begin{eqnarray}
\label{Qtau}
\lim_{\tau\to\infty} \frac{\log\left[p(\tau)\right]}{\tau}=-E^*,
\end{eqnarray}
where $E^*$ is the GS energy  
of a ``bosonic'' 
Hamiltonian restricted to the cavity and defined as follows:
$H^*=K^*+V^*$, where, for any $\bm{n},\bm{n}'\in \bar{\mathbb{F}}$, 
$K^*_{\bm{n},\bm{n}'}\defi |K_{\bm{n},\bm{n}'}|$, $K_{\bm{n},\bm{n}'}$ being the original kinetic matrix of the given system, 
and $V^*(\bm{n})\defi R(\bm{n})$, $R(\bm{n})$ being the (total) kinetic terms associated to $K_{\bm{n},\bm{n}'}$.
Note that, by construction, the kinetic terms of $H^*$ are given by
$R^*(\bm{n})=R_{\mathrm{in}}(\bm{n})$, whereas $R(\bm{n})=R_{\mathrm{in}}(\bm{n})+R_{\mathrm{out}}(\bm{n})$, with
$R_{\mathrm{in}}(\bm{n})$ and $R_{\mathrm{out}}(\bm{n})$ defined in Eqs. (\ref{Q24})-(\ref{Q24b}). 
Notice that: \textit{i)} if in the place of $V^*(\bm{n})=R(\bm{n})$ we had 
$V^{**}(\bm{n})=R_{\mathrm{in}}(\bm{n})$, the GS of the corresponding $H^{**}$ 
would be the uniform state with GS energy $E^{**}=0$,
\textit{ii)} for the same reason, and due to the fact that the cavity space is connected
to the reservoir (one of the hypothesis of our theorem), it is $E^*>0$. 
Interestingly, properties \textit{i)} and \textit{ii)} can be seen as
classical results of algebraic graph theory \cite{Godsil}, but they can also be derived 
immediately by using the EPR \cite{NOTE7}. 
We see here clearly the connection between the pdf $p(\tau)$ and the condition $E^*>0$. The fact
that the cavity be connected to the reservoir ensures that 
the probability to escape from the cavity in a finite time be finite, and that the decay of the pdf be exponential.

Now, the pdf given in Eq. (\ref{Qtau}) is an interesting element in statistical mechanics. Nevertheless,
it is not useful for our theorem: in our theorem we need a less difficult - but still exact - pdf as
the $p(\tau=x)$ given in Eq. (\ref{Q27tau}).
The perfect balance between weights and probabilities, condensed in Eq. (\ref{QLemma1}), 
is a feature of the EPR that other approximate probabilistic representations, based, \textit{e.g.},
in Trotter-Suzuki formula (see \textit{e.g.}, \cite{TRIVEDICEPERLEY}; see also \cite{OP3}), 
cannot satisfy, and a tentative to prove the theorem with them would fail. 
The exactness of the EPR is the crucial ingredient at the base of the first-order QPT's 
scenario stated in the theorem \cite{NOTE8}.

\section{Conclusions}
\label{conclusions}
Starting form an exact probabilistic representation of an arbitrary quantum dynamics \cite{BPDAJL}, 
a very general theorem on QPT phase transitions has been here derived.
The critical equation, which consists in the equality of the reservoir and cavity
energies $\widetilde{E}$ and $\bar{E}$ in the thermodynamic limit, under the condition
that the cavity space $\bar{\mathbb{F}}$ be infinitely diluted with weight
$\bar{p}=|\bar{\mathbb{F}}|/|{\mathbb{F}}|$ going to zero faster than $1/N$,
has a very appealing physical explanation as follows.
Once the system enters in the cavity-subspace $\bar{\mathbb{F}}$, if its
energy $\bar{E}$ is lower than the energy of the reservoir-subspace $\widetilde{E}$,
just due to the fact that $\bar{p}\to 0$, the system prefers not to leave the
cavity anymore, since, otherwise, once left the cavity, 
the probability for the system to find again the cavity would be very low and, during
its traveling in the Fock space, it would accumulate an higher energy $\widetilde{E}$.
On the other hand, when instead $\widetilde{E}<\bar{E}$,  
the system is no more forced to be in such a rare event
in the Fock space, so that, just due to the fact that $\bar{p}\to 0$,
the cavity is almost not visited and the system remains in the normal phase
with an energy given by $E\sim \widetilde{E}$ with a relative error of the order $\mathop{O}(\bar{p})$. 

We point out that this phase transition scenario picture takes
place in the Fock space and that the required infinite dilution $\bar{p}N\to 0$
has nothing to share with a material dilution. In fact, a necessary condition for the
theorem is that the potential levels $V$'s, as well as the kinetic terms $R$'s,
be extensive with $N$, a condition
that cannot be satisfied by an infinitesimal fraction of particles. 

We believe that our theorem, due to its universality, 
will have many important consequences and we
forecast to see in the future many interesting applications.

\section*{Acknowledgments}
This research was supported by 
SOCIALNET grant project 217141.
M.O. acknowledges C. Presilla for many important discussions
and the Department of Physics of Roma for partial support
and hospitality during the completion of this paper.

\appendix
\section{Proof of the Lemma (Eqs. (\ref{QLemma1})-(\ref{Qergo2}))}
\setcounter{section}{1}
\label{Appendice}

To prove Eq. (\ref{QLemma1}), let us now come back to the EPR. 
Recalling that after $N_t$ jumps the system visits $N_t+1$
configurations,
we can decompose $\E\left(\mathcal{M}^{[0,t)}\right)$ as
\begin{eqnarray}
\label{Q26}\fl
\E\left(\mathcal{M}^{[0,t)}_{\bm{n}_0}\right)&=& \sum_{r \in \Omega}\int_{0}^{'t}d\delta_0^{(r)}\ldots d\delta_{N_t-1}^{(r)}
p\left(\delta_0^{(r)},\ldots,\delta_{N_t-1}^{(r)}\right) p_{N_t}^{(r)}\nonumber \\
&& \times e^{\left[R_0^{(r)}-V_0^{(r)}\right]\delta_0}\lambda_1^{(r)} \cdots
e^{\left[R_{N_t}^{(r)}-V_{N_t}^{(r)}\right]\delta_{N_t}}\lambda_{N_t}^{(r)},   
\end{eqnarray}
where: $\Omega$ is the set of trajectories extracted with the probability $p_{N_t}^{(r)}$
(the Markov chain starting from $\bm{n}_0$), 
and - recalling Defs. (\ref{Lf}) -
$R_0^{(r)},\ldots,R_{N_t}^{(r)}$, $V_0^{(r)},\ldots,V_{N_t}^{(r)}$,
and $\lambda_1^{(r)},\ldots,\lambda_{N_t}^{(r)}$, are the sequences of the 
the kinetic terms, the potential values, and the signs, occurred along 
the $r$-th trajectory, respectively,
while $\delta_0^{(r)},\ldots,\delta_{N_t-1}^{(r)}$ are 
the sequences of the local living times
extracted with the density probability $p(\delta_0^{(r)},\ldots,\delta_{N_t-1}^{(r)})$ given by (see note \cite{NOTE1}) 
\begin{eqnarray}
\label{Q27}\fl
p(\delta_0^{(r)},\ldots,\delta_{N_t-1}^{(r)})=
R_0^{(r)}e^{-R_0^{(r)}\delta_0^{(r)}}\cdots R_{N_t-1}^{(r)}e^{-R_{N_t-1}^{(r)}\delta_{N_t-1}^{(r)}}\times e^{-R_{N_t}^{(r)}\delta_{N_t}^{(r)}}
\end{eqnarray}
and $\delta_{N_t}^{(r)}$ is the residual time given by
\begin{eqnarray}
\label{Q27b}
\delta_{N_t}^{(r)}=t-\left(\delta_0^{(r)},\ldots,\delta_{N_t-1}^{(r)}\right).
\end{eqnarray}
Finally, as in Sec. III, the symbol $\int_{0}^{'t}$ is a shorthand notation to indicate that the $N_t$ time intervals
$\delta_0^{(r)},\ldots,\delta_{N_t-1}^{(r)}$ are constrained by 
$0\leq \delta_0^{(r)}+\ldots+\delta_{N_t-1}^{(r)}\leq t$.
Note that, given the intervals $\delta_0^{(r)},\ldots,\delta_{N_t-1}$, 
the last factor $e^{-R_{N_t}^{(r)}\delta_{N_t}^{(r)}}$ in Eq. (\ref{Q27})
represents the probability (and not a pdf!) for the system not to jump during the residual time
interval $\delta_{N_t}^{(r)}$ (notice the absence of the factor $R_{N_t}^{(r)}$). 
Finally, let $\eta_1^{(r)},\ldots,\eta_{N_t}^{(r)}$
be the hopping terms occurred along the $r$-th trajectory.
According to the probability of the Markov chain, Eq. (\ref{Pf} ), the probability $p_{N_t}^{(r)}$
is given by
\begin{eqnarray}
\label{Q28}
p_{N_t}^{(r)}=\frac{\eta_1^{(r)}}{R_0^{(r)}}\cdots \frac{\eta_{N_t}^{(r)}}{R_{N_t-1}^{(r)}}.
\end{eqnarray}

Let now $\tau$ be some random time at which the system performs some task before $t$,
\textit{e.g.}, the first time at which the system enters in some target region of the Fock space.
Let us consider the expectation $\E\left(\mathcal{M}^{[0,\tau)}_{\bm{n}_0}\right)$.
We want now to decompose this expectation as we have done in (\ref{Q26}).
However here, unlike Eq. (\ref{Q26}), we have to pay attention to two new points: 
\textit{i)} $\tau$ is random variable taking values in $[0,t)$, \textit{ii)} by definition $\tau$ is also
the time at which a further jump takes place and sends the system in some region for the first time before $t$.
Given a trajectory starting from $\bm{n}_{0}$, 
let us indicate with $N_{\tau}$ the number of jumps performed by the system 
during the interval $[0,\tau)$. 
Note that now, the set of the $N_{\tau}$ time intervals 
$(\delta_0,\ldots,\delta_{N_{\tau}-1},\delta_{N_{\tau}})$, 
is a set of independent random variables being 
$\tau=\delta_0+\ldots+\delta_{N_{\tau}}$
a random variable whose pdf $p(\tau=x)$
is given by
\begin{eqnarray}
\label{Q27tau}
p(\tau=x)=R_{N_{x}}^{(r)}e^{-R_{N_{x}}^{(r)}\delta_{N_{x}}^{(r)}(x)},
\end{eqnarray}
where, given $x$, $\delta_{N_{x}}^{(r)}(x)$ is the residual time as given in Eq. (\ref{Q27b}):
\begin{eqnarray}
\label{Q27stoch}
\delta_{N_{x}}^{(r)}\defi x-\left(\delta_{0}^{(r)}+\ldots+\delta_{N_{x}-1}^{(r)}\right).
\end{eqnarray}
Note the difference with the previous case in which, given a trajectory with $N_t$ jumps,
the sum of the delta's in Eq. (\ref{Q26}) had to give the constant $t$.
Here we are in fact free to choose between one of the two sets of independent random variables 
to be integrated over the interval $[0,t)$:
$(\delta_0,\ldots,\delta_{N_{\tau}-1},\delta_{N_{\tau}})$ or 
$(\delta_0,\ldots,\delta_{N_{\tau}-1},\tau)$. 
We choose the latter, the Jacobian being 1. 
Furthermore, taking into account that, at the jump time $x$,
given $R(\bm{n}_{x^-})$ ($=R_{N_{x}}$) and $R_{\mathrm{out}}(\bm{n}_{x^-})$ ($=R_{\mathrm{out};N_{x}}$), 
the system can go to any of the $R_{\mathrm{out}}(\bm{n}_{x^-})$ configurations of the reservoir
with a total probability $R_{\mathrm{out}}(\bm{n}_{x^-})/R(\bm{n}_{x^-})$,
and by using Eqs. (\ref{Q27tau}) and (\ref{Q27stoch}), we arrive at
\begin{eqnarray}
\label{Q26stoch}\fl
\E\left(\mathcal{M}^{[0,\tau)}_{\bm{n}_0}\right)
&=& \int_{0}^{t}dx \sum_{r \in \Omega_x} \int_{0}^{'x} d\delta_0^{(r)}\ldots d\delta_{N_{x}-1}^{(r)}
p\left(\delta_0^{(r)},\ldots,\delta_{N_{x}-1}^{(r)}\right) 
p_{N_{x}}^{(r)} 
\nonumber \\
&& \times e^{\left[R_0^{(r)}-V_0^{(r)}\right]\delta_0}\lambda_1^{(r)} \cdots
e^{\left[R_{N_{x}}^{(r)}-V_{N_{x}}^{(r)}\right]\delta_{N_{x}}}\lambda_{N_{x}}^{(r)}~R_{\mathrm{out}}^{(r)}(\bm{n}_{x^-}),  
\end{eqnarray}
where $\Omega_x$ is the set of trajectories which, at the instant $x$, and not before, reach
the target region. In turn, from Eqs. (\ref{Q27}) and (\ref{Q28}), we see that the above expression
can be rewritten as
\begin{eqnarray}
\label{Q26stochb}\fl
\E\left(\mathcal{M}^{[0,\tau)}_{\bm{n}_0}\right)
&=& \int_{0}^{t}dx \sum_{r \in \Omega_x} \int_{0}^{'x} d\delta_0^{(r)}\ldots d\delta_{N_{x}-1}^{(r)}
p_{\mathrm{in}}\left(\delta_0^{(r)},\ldots,\delta_{N_{x}-1}^{(r)}\right) 
p_{\mathrm{in};N_{x}}^{(r)} 
\nonumber \\
&& \times e^{\left[R_{\mathrm{in};0}^{(r)}-V_0^{(r)}\right]\delta_0}\lambda_1^{(r)} \cdots
e^{\left[R_{\mathrm{in};N_{x}}^{(r)}-V_{N_{x}}^{(r)}\right]\delta_{N_{x}}}\lambda_{N_{x}}^{(r)}~R_{\mathrm{out}}^{(r)}(\bm{n}_{x^-}),  
\end{eqnarray}
where $p_{\mathrm{in}}\left(\delta_0^{(r)},\ldots,\delta_{N_{x}-1}^{(r)}\right)$ 
and $p_{\mathrm{in};N_{x}}^{(r)}$, are the pdf and the trajectory's probability, 
in which appear only the kinetic terms $(R_{\mathrm{in};0}\ldots,R_{\mathrm{in};N_{x}})$
and $(R_{\mathrm{in};0}\ldots,R_{\mathrm{in};N_{x}-1})$, respectively.
By observing that $p_{\mathrm{in};N_{x}}^{(r)}$ is normalized to 1
over the set $\Omega_x$, and by comparing with Eq. (\ref{Q26}) 
we see therefore that Eq. (\ref{Q26stochb}) can be read as
\begin{eqnarray}
\label{Q28stoch}
\E\left(\mathcal{M}^{[0,\tau)}_{\bm{n}_0}\right)
&=& \int_{0}^{t}dx ~\E\left(\mathcal{M}^{[0,x)}_{\mathrm{in};\bm{n}_0}R_{\mathrm{out}}(\bm{n}_{x^-})|~\tau=x\right).
\end{eqnarray}
The generalization to the case of $k$ random times is trivial.

Finally, to prove Eq. (\ref{Qergo1}), we observe that:
\textit{i)} the conditional expectation in the numerator of the lhs of Eq. (\ref{Qergo1}) is - by definition of
conditional expectation -
built with a probability normalized over the space of trajectories leaving the cavity at the time
$x_1$, \textit{ii)} as $x_1$ goes to $\infty$, 
the set of the weights $\bar{\mathcal{M}}^{[0,x_1)}$ (weighted trajectories) 
not included in the expectation of the numerator with respect to those of
the denominator, approaches 0. The proof of Eq. (\ref{Qergo2}) is equal.


\section*{References}
\begin{thebibliography}{99}


\bibitem{Sach} See for example: S. Sachdev \textit{Quantum Phase Transitions} (2000) Cambridge University Press.

\bibitem{OP5} Ostilli M and Presilla C, 
2006 \textit{J. Stat. Mech. P11012}


\bibitem{BPDAJL} Beccaria M, Presilla C, De Angelis G F and Jona-Lasinio G,
1999 \textit{Europhys. Lett.} {\bf 48}, 243

\bibitem{OP4} Ostilli M and Presilla C, 
2005 \textit{J. Stat. Mech.} P04007 

\bibitem{BM} A.\ J.\ Bray and M.\ A.\ Moore, J.\ Phys. C {\bf 13}, L469 (1980).

\bibitem{MPV} M.\ M\'ezard, G.\ Parisi, and M.\ A.\ Virasoro, {\it Spin Glass
    Theory and Beyond} (World Scientific, Singapore, 1984).

\bibitem{Obuchi} T.\ Obuchi, H.\ Nishimori, and D.\ Sherrington,
  J.\ Phys.\ Soc.\ Jpn.\ {\bf 76} 054002 (2007).

\bibitem{SHY} See for example: Shiryayev A N, 
1984 \textit{Probability} (New York: Springer)

\bibitem{PSPIN} B.\ Derrida, Phys.\ Rev.\ Lett.\ {\bf 45}, 79
  (1980); D.\ J.\ Gross and M.\ M\'ezard, Nucl.\ Phys.\ B {\bf 240}, 431
  (1984).

\bibitem{Krzakala} T. J$\mathrm{\ddot{o}}$rg, F. Krzakala, J. Kurchan, and A. C. Maggs,
Phys. Rev. Lett. 101, 147204 (2008).




\bibitem{OP6} Ostilli M and Presilla C, 
in preparation.


\bibitem{Godsil} 
Godsil, C., Royle, G. \textit{Algebraic graph theory}, in Graduate Texts in Mathematics, 
vol. 207 (Springer-Verlag, New York 2001). 


\bibitem{TRIVEDICEPERLEY} See for example: N. Trivedi and D. M. Ceperley, Phys. Rev. B \textbf{41}, 4552 (1990).

\bibitem{OP3} Ostilli M and Presilla C, 
2004 \textit{J. Phys.} A \textbf{38}, 405.

\bibitem{NOTE}
However, in general, it is not necessary to require that 
$H$ be the Fock representation of some true Hamiltonian operator~$\hat{H}$. 

\bibitem{NOTE0}
We point out that in the proof of the theorem there appears only the invariant measure $\bar{\pi}$
associated to the cavity space, and not the ``flat'' measure $\bar{p}$.
It is possible therefore to formulate the theorem without requiring that be
$\bar{p}N\to 0$, but only $\bar{\pi}N\to 0$. This more abstract formulation is a bit
more general and allows to consider systems for which not all the kinetic terms $K(\bm{n})$
be extensive, but only a finite fraction of them, or even systems having, in the Fock space,
a finite fraction of kinetic terms going to 0 in the thermodynamic limit.

\bibitem{NOTE1}
Note in general that, the pdf defined as $p(s)=\exp (-sR)R$, is normalized to 1 only in the interval $[0,+\infty)$,
whereas the correct normalized pdf for the finite interval $[0,t)$ is $p(s)=\exp (-sR)R/(1-\exp(-tR))$.
However, since we are interested only in the limit $t\to\infty$, from now on, throughout the paper, 
to avoid too cumbersome equations,
the effective approximation $R\simeq R/(1-\exp(-tR))$ will be left as understood.

\bibitem{NOTE2}
Sometimes, to stress the fact that a configuration $\bm{n}$ belongs to the 
reservoir or cavity spaces, we will make use of the symbol $\widetilde{\bm{n}}$
or $\bar{\bm{n}}$, respectively.

\bibitem{NOTE3} Notice the null role of the random variables $t_1,\ldots,t_k$ in the limit $t\to\infty$.

\bibitem{NOTE4}
The precise quantity giving exactly the average number of jumps would imply the knowledge
of $|E^{(0)}|$, where $E^{(0)}$ is the GS energy of the system with $V\equiv 0$. 

\bibitem{NOTE5} 
It is possible to reach more formally this result through the decomposition $Q_t(k)=\sum_{N=k}^\infty Q(k|N)P_t(N)$,
where $P_t(N)$ is the probability to have a total of $N$ jumps during the interval $[0,t)$ and $Q(k|N)=\mathcal{P}_N(k)$ 
is the probability for the Markov chain to visits $k$ times the cavity.
Then, by using for $P_t(N)\simeq (\langle R\rangle t)^N/N!\exp(-\langle R\rangle t)$ 
(see \cite{NOTE4}),   
we see that, for $t\gg kt_c$ $Q_t(k)\to(\pi(\bm{n}))^k$.

\bibitem{NOTE6}
We point out that, in the remind of the proof of the theorem,
just to avoid an excessive and cumbersome formalism, 
we will leave the symbols $t_1,\ldots,t_k$ as random variables
which, for the final result, do not play any role.
The skeptical reader may rewrite all the equations with fixed values $t_1=y_1,\ldots,t_k=y_k$, and,
only at the end of the calculation, to integrate the corresponding conditional expectation multiplied
by the pdf $p(t_1=y_1,\ldots,t_k=y_k)$. Since the conditional expectation, at the leading order,
does not depend on the variables $y_1,\ldots,y_k$, it is clear that, at the leading order, 
carrying on the calculation with the expectation conditioned by the $t$'s,
is completely useless. We stress again that, in our decomposition, 
the role of the random variables $t$'s is very different from
the role of the random variables $\tau$'s, and we need to use the Lemma only for the latter.

\bibitem{NOTE7} In fact, the property \textit{i)} comes immediately from the EPR
and the definition of the functional $\mathcal{M}^{[0,t)}$ which, in the
case of the system with Hamiltonian $H^{**}$, becomes simply $\mathcal{M}^{**;[0,t)}\equiv 1$.
Concerning the property \textit{ii)}, since the cavity is connected to the reservoir,
there are configurations of the cavity such that $R^*_{\mathrm{out}}>0$, so that, since $V^*\geq R^*$, it must be
also $E^*>E^{**}=0$. 


\bibitem{NOTE8} C. Presilla and G. Jona-Lasinio, at the time in which the EPR was derived, 
emphasized the exactness of the EPR, and hypothesized possible advantages not
featured by other approximate representations.

\end{thebibliography}
\end{document}